\documentclass[a4paper, pre, twocolumn,superscriptaddress,floatfix,amssymb,showpacs]{revtex4-1}
\usepackage{etex}
\usepackage{enumerate}
\usepackage{times}
\usepackage{graphicx}
\usepackage{amsmath}
\usepackage{xcolor}
\usepackage{natbib}
\usepackage{xspace}
\usepackage{tabularx}

\usepackage[pdfpagemode=None,colorlinks=true,urlcolor=black,%
linkcolor=blue,citecolor=blue,pdfstartview=FitH]{hyperref}
\graphicspath{{fig/}} 
\usepackage{soul}

\definecolor{greenx}{RGB}{0, 175, 124}

\definecolor{purplex}{RGB}{132, 10, 134}

\usepackage{ulem}

\newcommand{\ie}{i.e., }
\newcommand{\trans}{\textit{trans}\xspace}

\newcommand{\ka}{$\kappa$\xspace}
\newcommand{\No}{$N_0$\xspace}
\newcommand{\ro}{$\rho_0$\xspace}
\DeclareMathOperator{\sign}{sign}

\begin{document}

\title{Rigidity-induced scale invariance in polymer ejection from capsid}

\author{R. P. Linna}
\email{Author to whom correspondence should be addressed: riku.linna@aalto.fi}
\author{P. M. Suhonen}
\author{J. Piili}
\affiliation{Department of Computer Science, Aalto University, P.O. Box 15400, FI-00076 Aalto, Finland}

\begin{abstract}
While the dynamics of a fully flexible polymer ejecting a capsid through a nanopore has been extensively studied, the ejection dynamics of semiflexible polymers has not been properly characterized. Here we report results from simulations of ejection dynamics of semiflexible polymers ejecting from spherical capsids. Ejections start from strongly confined polymer conformations of constant initial monomer density. We find that, unlike for fully flexible polymers, for semiflexible polymers the force measured at the pore does not show a direct relation to the instantaneous ejection velocity. The cumulative waiting time $t(s)$, that is, the time at which a monomer $s$ exits the capsid the last time, shows a clear change when increasing the polymer rigidity $\kappa$. Major part of an ejecting polymer is driven out of the capsid by internal pressure. At the final stage the polymer escapes the capsid by diffusion. For the driven part there is a cross-over from essentially exponential growth of $t$ with $s$ of the fully flexible polymers to a scale-invariant form. In addition, a clear dependence of $t$ on $N_0$ was found. These findings combined give the dependence $t(s) \propto N_0^{0.55} s^{1.33}$ for the strongly rigid polymers. This cross-over in dynamics where $\kappa$ acts as a control parameter is reminiscent of a phase transition. This analogy is further enhanced by our finding a perfect data collapse of $t$ for polymers of different $N_0$ and any constant $\kappa$.

\end{abstract}


\maketitle
\section{Introduction}\label{sec:intro}
Due to advances in biosciences the behavior of macromolecules such as DNA and proteins has been under intensive study over the last two decades. Living cells rely heavily on these biopolymers, which are  transported through membranes via various processes. The escape of DNA and RNA from viral capsids is one of the most important such  transportation processes. Viruses work by injecting their DNA or RNA to the host cell. A multitude of viruses store their genome inside a more or less spherical shell and eject it to a host cell through a pore or a syringe. Some well-known examples are the T4 phage, the T7 phage and the $\lambda$ phage. 

Neither biological nor theoretical perspectives of the mechanisms of packaging, activation, and release of DNA from viral capsids are thoroughly understood~\cite{molineux_popping,grayson}. The motivation to understand fundamentals of this process is further enhanced by its potential applications in drug delivery and gene therapy~\cite{glasgow}. Recently significant advancement was made in artificially storing data in DNA~\cite{erlich_dna_fountain_storage}. This could open up possibilities for storing human-generated data into capsids from where it could be retrieved by using sequencing techniques.

The field of biopolymer dynamics has inspired physicists, since it provides complex and theoretically challenging problems to study. Accordingly, also capsid ejection of polymers has been the subject of a number of theoretical treatments~\cite{muthukumar1, cacciuto2, grosberg_stafy, sakaue_polymer_chains, sakaue_polymer_decompression}. These theoretical treatments have been accompanied by a multitude of studies using computer simulations~\cite{ali_shape_matters, ali_capsid_solvent_quality, lawati_capsid_tail, riku_dynamics_of_ejection, piili_capsid, piili_capsid2, mahalik_langevin_phage}. These studies have addressed the ejection of fully flexible polymers and as such provide important information on the fundamentals of ejection dynamics. The results obtained from them have relevance to viruses containing RNA, such as polio virus~\cite{alberts_molecular}, or single stranded DNA that eject their genome through a pore. 

Semiflexible polymers have received considerably less attention in both theoretical and computational studies of capsid ejection~\cite{marenduzzo_topological_dna_ejection, lawati_capsid_tail, mahalik_langevin_phage, marenduzzo_dna_interactions_knot, ali_capsid_solvent_quality, mahalik_langevin_phage}. The double stranded DNA (dsDNA) shows resistance to bending, which in some cases of polymer dynamics does not change the characteristics appreciably from those of the fully flexible chain~\cite{linna_dna}. However, taking into account the bending rigidity of the dsDNA is very important in cases where confined geometries are involved. In dsDNA ejection it is essential, since the size of viral capsids is typically of the same order as dsDNA's persistence length ($\sim 50$ nm \cite{manning2006persistence}).

Here, we study the ejection of semiflexible polymer chains from a spherical capsid. Computer simulations are performed using our implementations of stochastic rotation dynamics and Langevin dynamics. It has been observed that in vivo, the temporal aspects of the process are governed by long pauses throughout the ejection~\cite{van_valen_hershey-chase}. Mahalik et al. were able to reproduce this in simulations of semiflexible polymers~\cite{mahalik_langevin_phage}. They also explained that the ejection speed is directly related to the angle of incidence to the pore. While the friction of the pore has an indisputable effect on translocation in real biological systems, in this paper, which is the third in the series of papers investigating capsid ejection for different relevant settings, we focus on the underlying general ejection process. In order to do this we have reduced the pore friction by implementing a pore that has a smooth geometry.

Using our general simulation model, we will study the ejection process for polymers of different bending rigidity. In our previous studies on capsid ejection of flexible polymers we showed that the ejection dynamics is governed by the force the polymer beads inside the capsid exert on the bead at the pore entrance~\cite{piili_capsid, piili_capsid2}. As will be seen, this result cannot be generalized to semiflexible polymer chains.

The paper is organized as follows. In Section~\ref{mm} we describe the used computational models and methods. Results are presented in Section~\ref{sec:res}. Here, we first give account of how polymer packaging rate affects ejection in \ref{sec:res:initConf}, after which we evaluate the contribution to dynamics coming from the {\it trans} side, i.e. outside the capsid. Measurements of ejection times and pore force are reported and analyzed in~\ref{sec:res:pore} and~\ref{sec:poreforce}, respectively. The central results and analysis of the cross-over of ejection dynamics with increasing polymer rigidity is given in \ref{sec:scaling}. In Section~\ref{sec:sc_functions} we show that the measurement data describing ejection dynamics of polymers of same rigidity but different length can be made to collapse onto a single curve. Results and conclusions are summarized Section~\ref{sec:con}. The main result of the present study is the observation that ejection dynamics enters a scale-invariant regime as polymer rigidity is increased.

\section{Models and methods}
\label{mm}

\begin{figure}
\centering
\includegraphics[width=\linewidth]{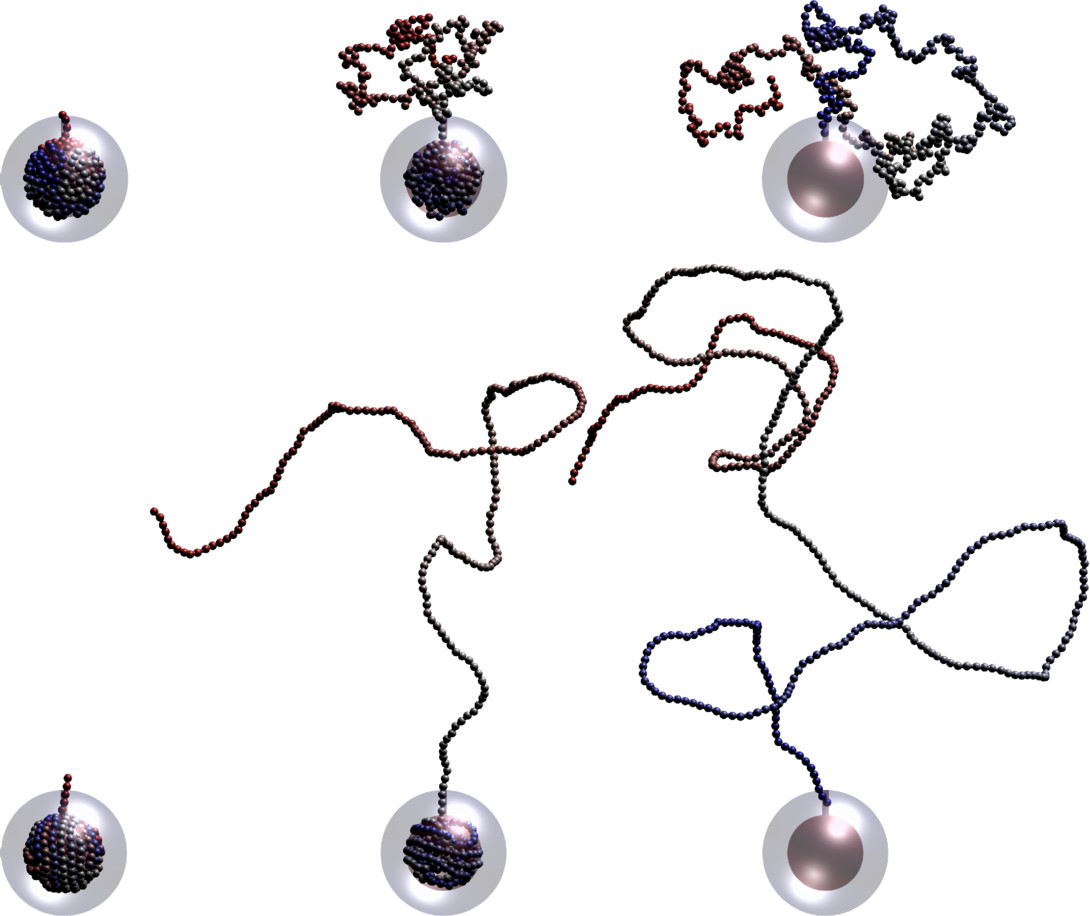}
\caption{(Color online) Snapshots of capsid ejection simulations at the start of the simulation (left), halfway through translocation (middle), and in the end (right). Top row: flexible chain ($\kappa=0$). Bottom row: semiflexible chain ($\kappa=20$). Polymer length $N_0=400$ and initial monomer density $\rho_0=0.826$.}
\label{fig:capsidPicture}
\end{figure}

Here we present the computational model used in this study. For the polymer we use a coarse grained beads-and-springs model with excluded volume interactions. The polymer is immersed in solvent modeled via stochastic rotation dynamics (SRD) which can be used to simulate hydrodynamical interactions. In addition to SRD, we use the computationally more effective Langevin dynamics to create initial conformations and measure pore force. The capsid is modeled as rigid walls with slip and no-slip boundary conditions for the polymer and the solvent, respectively. The simulation geometry and snapshots of an ejecting polymer are depicted in Fig.~\ref{fig:capsidPicture} for polymers of different persistence lengths.

\subsection{The polymer model}
The polymer is a chain of point-like beads where each subsequent pair of beads is connected with the finitely extensible nonlinear elastic (FENE) potential
\begin{align}\label{equ:fene}
U_{\rm F} = - \frac{K}{2} r_{\rm max}^2 \ln{\left( 1 - \left(\frac{r}{r_{\rm max}} \right)^2 \right)}\;,\; r < r_{\rm max}.
\end{align}
Here $r$ is the distance between the two subsequent bead and $K$ and $r_{\rm max}$ are potential parameters. Between all polymer beads we have the shifted and truncated Lennard-Jones (LJ) potential that accounts for the excluded volume interactions
\begin{align}\label{equ:lj}
U_{\rm LJ} = \left\lbrace \begin{array}{ccl}
		4.8 \epsilon \left[\left(\frac{\sigma}{r_{ij}}\right)^{12} - \left(\frac{\sigma}{r_{ij}}\right)^6\ \right] + 1.2\epsilon &,& r_{ij} \leq \sqrt[6]{2}\sigma\\
		0		&,& r_{ij} > \sqrt[6]{2}\sigma
		\end{array},
 \right.
\end{align}
Here $r_{ij}$ is the distance between beads with indices $i$ and $j$. The potential is non-attractive everywhere and therefore models a good solvent. The parameters for the polymer potentials are chosen as $\sigma = 1.0$, $\epsilon = 1.0$, $K = 30/\sigma^2$, and $r_{\rm max} = 1.5\sigma$ in reduced units.

To model bending rigidity we use the total bending potential of the form
\begin{align}
U_{\rm bend} = -\kappa \sum_{i} \left( \mathbf{r}_{i+1} - \mathbf{r}_{i} \right)\cdot\left( \mathbf{r}_{i} - \mathbf{r}_{i-1} \right),
\end{align}
where the parameter \ka defines the strength of the bending potential so that persistence length grows with $\kappa$. In other words the potential resists bending and has its minimum when the polymer is straight.

\subsection{The solvent and polymer dynamics}
We use stochastic rotation dynamics (SRD) to model the solvent in which the polymer is immersed (sometimes referred to as multi-particle collision dynamics)~\cite{malevanets_orig, malevanets_mesoscopic,gompper_srd}. SRD models full hydrodynamic interactions as well as thermal fluctuations. It is also computationally effective and allows for switching off hydrodynamic interactions when needed.

The SRD solvent consists of point-like particles that are propagated in two alternating steps: the \textit{collision step} and the \textit{streaming step}. In the \textit{streaming step} the particles' positions are updated using the update rule
\begin{align}
\mathbf{r}_i(t + \Delta t) = \mathbf{r}_i(t) + \mathbf{v}_i(t)\Delta t,
\end{align}
where $\mathbf{r}_i$ and $\mathbf{v}_i$ are the position and velocity of the particle $i$, respectively. $\Delta t$ is the SRD time step. For the \textit{collision step}, the particles are divided into a grid of cubic cells of edge length $1$. In each cell the velocities of the particles are updated via the formula
\begin{align}\label{equ:srd_step}
\mathbf{v}_i(t+\Delta t) = \mathbf{v}_{\rm cm}(t) + \Omega \left[ \mathbf{v}_i(t) - \mathbf{v}_{\rm cm}(t) \right],
\end{align}
where $\mathbf{v}_{\rm cm}$ are the center of mass of the cell and $\Omega$ is a rotation matrix unique for the cell. The rotation matrix $\Omega$ has a rotation angle $3\pi/4$ and its rotation axis is randomly chosen for each cell each time step. The rotation angle defines the viscosity of the solvent. The collision step conserves energy and momentum in each cell. To preserve Galilean invariance the collision grid is shifted randomly at each step~\cite{ihle_galilean}. Between collision steps the random parts of the velocities of the solvent particles are scaled such that the solvent stays at the constant temperature of $kT=1.0$~\cite{frenkel_moldy}.

The polymer is integrated in time using the velocity Verlet algorithm with time step $\delta t=0.0002$~\cite{swope_velocity_verlet, frenkel_moldy}. Such a small time step was chosen to avoid cumulation of numerical errors due to the high monomer densities inside the capsid. For SRD we use the time step of $\Delta t = 0.5$. The polymer is coupled to the SRD solvent via the collision step where the polymer particles and solvent particles are treated similarly. Velocity Verlet and SRD algorithms take turns such that after $\Delta t/\delta t = 2500$ velocity Verlet steps a single SRD step is taken. The solvent particles have mass $m_s=4$ and the polymer particles have mass $m_b=16$.

One of the benefits of SRD is that we can switch hydrodynamics off in order to better understand its effects. This is achieved by randomly interchanging the velocities of each particle every step so that the hydrodynamic correlations vanish.

In addition to SRD, we use Langevin dynamics (LD) to model solvent in force measurements and polymer packaging due to its superior computational efficiency. In~\cite{piili_capsid2} we showed that LD and SRD without hydrodynamics yield similar results. LD follows the Langevin equation
\begin{align}\label{equ:langevin}
m_b\frac{{\rm d}\mathbf{v}_i}{{\rm d} t}(t) = -\xi m_b \mathbf{v}_i(t) + \boldsymbol{\eta}_i(t) + \mathbf{f}_i(t),
\end{align}
where $\xi$ is the friction constant $\boldsymbol{\eta}_i(t)$ is a random force exerted on the bead $i$ and $\mathbf{f}_i(t)$ includes all the external forces exerted on the bead $i$. We integrate the Langevin equation in time using Ermak's implementation~\cite{ermak}. The LD parameters were matched to those of SRD according to~\cite{piili_capsid2}. The time step in LD simulations was chosen to be $\delta t=0.001$. In LD hydrodynamics is not included, but the solvent constitutes a Brownian heat bath.

\subsection{The simulation geometry}
The simulation geometry is depicted in Fig.~\ref{fig:capsidPicture}. The polymer ejects from the inside (\ie the \textit{cis} side) to the outside (\ie the \textit{trans} side) of the spherical capsid shell through a narrow pore. In the initial state before ejection there are $N_0$ beads inside the capsid and a tail of 4 beads in the pore and outside the capsid. Therefore, the total length of the polymer is $N_0+4$. The thickness of the capsid shell is 3. The radius $R_0$ of the inner shell of the capsid depends on the chosen initial monomer density \ro and \No via
\begin{align}
 \rho_0 = \frac{N_0}{\frac43 \pi (R_0+\zeta)^3},
\label{eq:density}
\end{align}
where $\zeta=0.3$ is added to the capsid radius to account for the portion of the beads' volume residing outside the capsid inner wall. After this correction the measured pore forces collapse when plotted as a function of density for fully flexible polymers~\cite{piili_capsid2} (there defined as effective density). In other words, for fully flexible polymers of different lengths the force measured at the pore is approximately the same when the monomer density inside the capsid $\rho=3N/(4\pi(R_0+\zeta)^3)$  is the same. $N$ is the number of monomers inside the capsid. Notice that in some publications volume fraction $\phi_0 = 4/3\pi\rho_0$ is used instead. Also the beads have often a hard sphere potential instead of the soft sphere potential used here. The hard sphere potential is not possible in simulations performing real dynamics at least for densities as high as used here. Accordingly, the values for densities in studies by different authors are not directly comparable. Unless otherwise noted, in the present study the initial monomer density $\rho_0 = 0.75$

\begin{figure}
\includegraphics[width=0.49\linewidth]{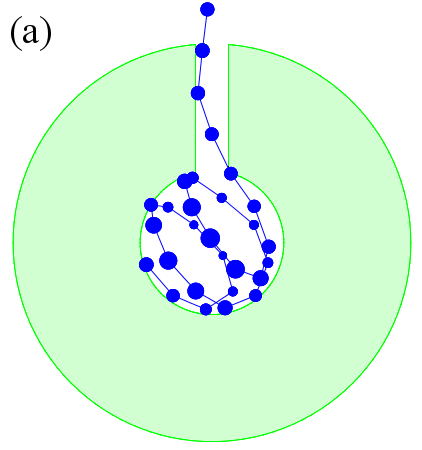}
\includegraphics[width=0.49\linewidth]{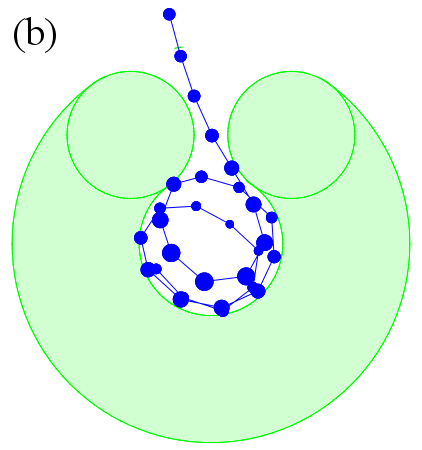}
\caption{Two dimensional projection of the capsid geometry with different pores. Initial configuration of polymer of length $N_0=25$ in capsids of initial monomer density $\rho_0=0.75$ with (a) cylinder pore (b) torus pore. (Monomer radius depicts distance from the viewer).}
\label{fig:capsidPores}
\end{figure}

To model the pore, we use a torus shaped opening in the capsid shell as depicted in Fig.~\ref{fig:capsidPores}~(b). We also compare this pore model to the more conservative cylinder pore, cf. Fig.~\ref{fig:capsidPores}~(a). It is evident that the form of the pore has an effect on ejection dynamics. This is especially true for polymers exhibiting high bending resistance, since they must straighten considerably in the pore region to eject. This effect has been previously seen in Ref.~\cite{mahalik_langevin_phage} where the approach angle to the pore during ejection was observed to determine the momentary ejection rate. We use the smoother toroidal pore to attenuate these effects so as to make more general inferences of the escape process.

The whole capsid geometry is created using constructive solid geometry technique~\cite{wyvill_csg}, which we have implemented for use in our simulations. In the simulations using a cylinder pore, the inner sphere and the pore (cylinder) are subtracted from the outer sphere [${\rm Capsid} = {\rm Sphere}_{\rm out} \setminus {\rm Sphere}_{\rm in} \setminus {\rm Cylinder}_{\rm pore}$]. In the torus pore model the geometry is a bit more complicated. The inner part of the torus is obtained by an intersection with a cone, after which a union is taken of this intersection and the sphere shell cut by a similar cone [${\rm Capsid} = ({\rm Sphere}_{\rm out} \setminus {\rm Sphere}_{\rm in} \setminus {\rm Cone}) \cup ( {\rm Torus} \cap {\rm Cone} )$]. The method tracks intersections with the particles' trajectories and capsid walls. The interaction between the capsid and the polymer is implemented using slip boundary conditions. For the solvent we use no-slip boundary conditions, which make the solvent velocity at the boundary disappear~\cite{lamura_srd_poseuille}. The pore radius is $0.4$ for the polymer and $0.8$ for the solvent. The larger pore for the solvent allows for smoother fluid flow in the pore region.

\subsection{Creating initial polymer conformations}
\label{sec:creatingInitialConf}
Experiments indicate that DNA is heavily organized inside capsids~\cite{cerritelli_encapsidated}. This raises the question of how the polymers should be packaged for ejection simulations. For semiflexible polymer chains the method of packaging has been observed to have a tremendous impact on the resulting conformations ~\cite{muthukumar2,rapaport_polymer_packaging,stoop_packing_elastic_wires}. Also in Ref.~\cite{mahalik_langevin_phage} it was observed that the magnitude of the packaging force affects the ejection rate of the polymers. Knots induced by packaging are also known to slow down the ejection~\cite{matthews_knot_ejection, marenduzzo_topological_dna_ejection}.

As in the present study the focus  is on ejection dynamics for polymers of different lengths, we paid special attention in creating initial conformations in a consistent way such that the ejections with different parameters can be compared. We package the polymer inside the capsid one bead at a time and let the polymer relax in between the packaging steps. A resemblant packaging scheme is used in Ref.~\cite{locker_viral_packing} where it is justified by the way ATP driven packaging might occur in viral capsids.

Before starting the packaging, we place the last bead to eject in the middle of the pore and generate the rest of the polymer on the \textit{trans} side as a self-avoiding random walk. After this, we drag beads one at a time from outside into the pore with a harmonic force $\mathbf{f}_{\rm i,drag}=100 \cdot(\mathbf{r}_{\rm mid pore}-\mathbf{r}_{i})$. When a bead arrives to the middle of the pore, we pause the packaging for a time $t_{\rm eq}$ by fixing the dragged bead in place with different harmonic force $\mathbf{f}_{\rm i,fix}=1000 \cdot(\mathbf{r}_{\rm mid pore}-\mathbf{r}_{i})$. This allows the polymer segment inside the capsid to relax. After the time $t_{\rm eq}$ has passed, the bead is guided into the capsid with a force of magnitude $100$ and next bead is dragged into the pore. With a long enough time $t_{\rm eq}$ between injecting new beads, we are able to generate initial configurations that are close to equilibrium. During packaging the polymer dynamics is propagated in time using LD instead of SRD. This is done due SRD simulations being computationally intensive for large intermediate equilibration times $t_{\rm eq}$. Using LD is justified here, since hydrodynamics has no effect on this slow packaging dynamics.

As we will show in Section~\ref{sec:res:initConf}, the ejection times depend on the equilibration time $t_{\rm eq}$ used during the packing process. Therefore to create packed configurations for ejection simulations, we choose a $t_{\rm eq}$ large enough so that the ejection times seem to have converged. This way we ensure that polymers of different lengths are packaged in a consistent way.

\subsection{Pore force}
\label{sec:force}

For fully flexible polymers waiting times $t_w(s)$ (\ie the time it takes for bead $s$ to eject the capsid the last time after bead $s-1$ has ejected the capsid the last time) reflect the force $f(s)$ measured at the pore~\cite{piili_capsid,piili_capsid2}. To see if this is the case also for semiflexible polymers we measure $f(s)$. The measurement is performed in the same way as polymer packaging, only in the opposite direction. We start the measurements from the same initial conformations as the ejection simulations. Instead of fully releasing the polymer, we first hold the bead $s=1$ at the pore center by a harmonic force for a time $t_{\rm eq} = 2000$. After this we measure the force average for $t_{\rm mt}=2000$ time units while still holding the polymer in place. Next, the bead $s=2$ is pulled to the pore using a harmonic force and a similar equilibration and averaging is performed as for the first bead. We repeat this process for all beads and finally obtain the measured force for each $s$. 

The measured $f(s)$ is the absolute value of the force having the sign of the $z$-component of the force. ($z$- axis points outward and passes through the middle of the pore.) 
\begin{align}
f(s) = \frac{1}{J}\sum_{j=1}^J \sign{(\mathbf{f}_{j,z}(s))}|\mathbf{f}_j(s)|,
\end{align}
where $j$ is the index of an individual simulation, and $\mathbf{f}_j(s)$ is the time-average of the force over the measurement time $t_{\rm mt}$ in simulation $j$. $J$ is the number of simulations. The sign function is used to cope with occasionally inward-pointing force at large $s$, where force due to pressure inside the capsid has decayed to a low magnitude. The absolute value describes the ejection force better than just the $z$-component, because polymers of higher bending rigidity usually eject in a direction, which is at an angle to the pore axis $z$.

\subsection{About persistence length and unit mapping}
\label{sec:pers_lenth}

\begin{table}
\caption{Average bond length $b$ and persistence length $\lambda_p$ for different values of bending potential parameter $\kappa$. Measured for polymer of length $N_0=25$ in free solvent.}
\label{tab:kappaVsPersistence}
\begin{tabular}{c|c|c|c|}
\ka & $b$ & $\lambda_p$\\\hline
0 & 0.980 & 0.889 \\
5 & 0.988 & 4.35 \\
10 & 1.00 & 9.41 \\
15 & 1.01 & 15.3 \\
20 & 1.03 & 21.9 \\
\end{tabular}
\end{table}

Previously, we have studied the ejection of fully flexible  polymers~\cite{riku_dynamics_of_ejection, piili_capsid, piili_capsid2}. The main interest has been in understanding general aspects of the process by minimizing the details involved. While fully flexible polymers are suited for describing single stranded DNA and RNA, a vast number of biologically relevant capsid ejections involve double stranded DNA. In such systems, bending rigidity cannot be omitted. Here we compare the polymer model used here to the real DNA using a simple unit mapping.

For polymers in our model, the Lennard-Jones potential implies the width of $\sim 1$ simulation units. Hence, a single simulation length unit can be taken to correspond to the width of dsDNA, which is about $2$ nm. The persistence length $\lambda_p$ can be adjusted via the bending parameter \ka. By simulating a polymer in free solvent and measuring the angle $\theta$ between adjacent bonds and bond length $b$, we can compute the persistence length using the relation~\cite{hsu2012stretching}
\begin{align}
 \lambda_p = \frac{b}{- \ln\langle\cos\theta\rangle}.
\end{align}
The values of $\lambda_p$ (in simulation units) for a few different values of \ka are presented in Table~\ref{tab:kappaVsPersistence}. The persistence length of real dsDNA is about 50 nm, and therefore we can conclude that $\kappa = 20$ best describes the properties of the dsDNA in our simulations. However, to obtain the best possible understanding of the effects of bending rigidity, we perform simulations with multiple values of \ka.

\section{Results}\label{sec:res}
Here, we present the results obtained from our SRD and LD simulations. All the presented quantities are arithmetic means over at least $45$ independent simulations. In some cases more simulations were run in order to obtain better statistics.

\subsection{Packaging rate}
\label{sec:res:initConf}
\begin{figure}
\centering
\includegraphics[width=0.49\linewidth]{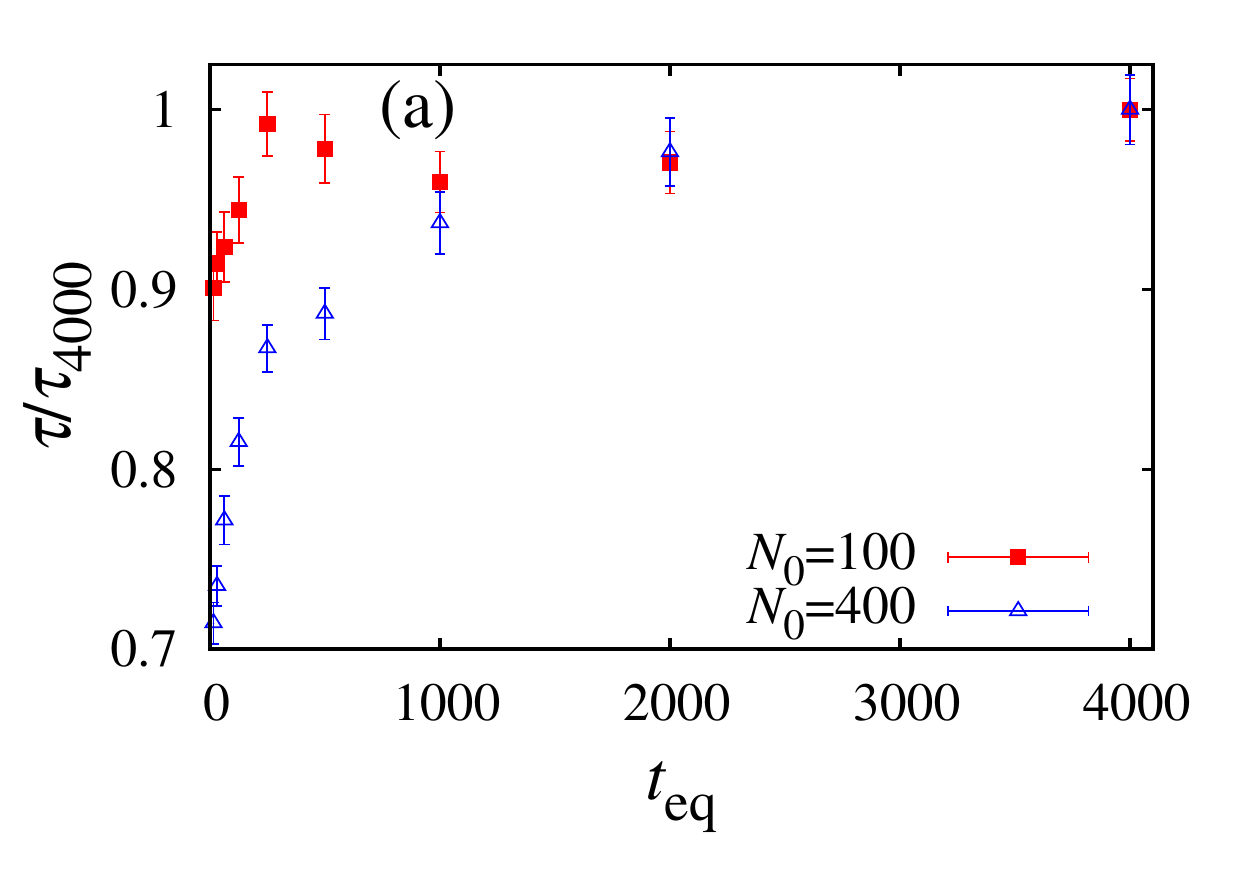}
 \includegraphics[width=0.49\linewidth]{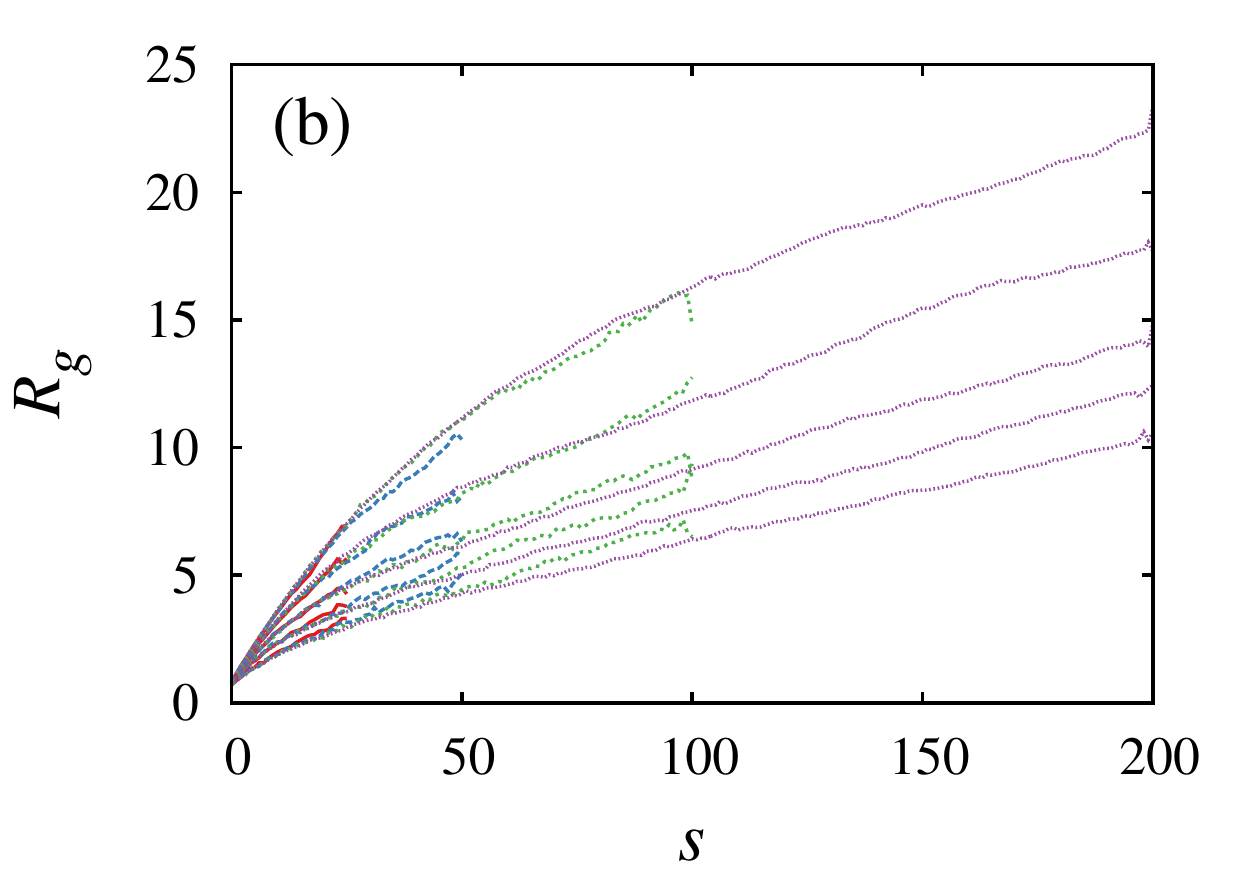}
\caption{(Color online) (a) Ejection times when different equilibration times are used $\tau(t_{\rm eq})$ normalized by $\tau(t_{\rm eq}=4000) \equiv \tau_{4000}$. LD and the cylinder pore model. The initial densities are $\rho_0=0.743$ and $0.826$ for $N_0=100$ and $400$, respectively, and $\kappa = 20$. Error bars depict the standard deviations of the mean. (b)  Radius of gyration $R_g$ on the \trans side during ejection for polymers of different length $N_0$. From top down $\kappa=20,10,5,2.5,$ and $0$.}
\label{fig:tauVsPacktime}
\end{figure}

As discussed in Section~\ref{sec:creatingInitialConf}, it is expected that when the polymer has bending rigidity, packaging affects the obtained initial polymer conformations and ejection times~\cite{mahalik_langevin_phage}. In order to define a consistent packaging method, we measured the ejection time $\tau$ as a function of the time $t_{\rm eq}$ used to equilibrate the polymer for each $s$ during the packaging, see Section~\ref{sec:creatingInitialConf}. This is shown in Fig.~\ref{fig:tauVsPacktime}~(a) for polymers of lengths $N_0 = 100$ and $200$ and for $\kappa=20$, simulated using LD and the cylinder pore.

We found that the longer the polymer is allowed to equilibrate during packaging, the slower becomes the ejection. This is in agreement with the observation that polymers that were packaged using a higher force eject faster~\cite{mahalik_langevin_phage}. The faster ejection of more rapidly packaged polymers occurs due to the polymer not having sufficient time to relax to an equilibrium conformation. After the polymer is packaged its conformational changes are extremely slow due to high energy barriers and therefore it remains in its local energy minimum. The local energy minima are deeper for a conformation that is closer to equilibrium, which slows down ejection.

Evidently, during packaging longer $t_{\rm eq}$ is required for longer polymers, as seen in Fig.~\ref{fig:tauVsPacktime}~(a).  Using $t_{\rm eq}=2000$ is seen to result in sufficiently equilibrated conformations at least for $N_0 \leq 200$. In addition, the effect of $t_{\rm eq}$ on ejection times was seen to diminish for $\kappa = 0$ (not shown). Consequently, $t_{\rm eq}=2000$ is a judicious choice for our simulations.

\subsection{Friction on the \trans side }
\label{sec:rg}

To guide the analysis of the ejection dynamics we first evaluate the contribution from the {\it trans} side friction, which has been shown to slow down ejection of semiflexible polymers~\cite{lawati_capsid_tail}. Fig.~\ref{fig:tauVsPacktime}~(b) shows the measured radius of gyration $R_g$ of the ejected polymer segment. On each point on the $s$-axis there is an identical number of beads on the {\it trans} side. As expected, $R_g$ increases with \ka. It is seen that $R_g$ grows fairly similarly with $s$ for different $N_0$, particularly for rigid polymers. Therefore, it is reasonable to expect that for all $s$ the \trans-side friction is very similar for semiflexible polymers of different length. On the other hand, for polymers of low rigidity the \trans-side friction is small, as was shown for $\kappa=0$~\cite{piili_capsid2}. Accordingly, the {\it trans} side friction does not cause differences between polymers of different length and the obtained characteristics are explainable by friction and conformational changes on the {\it cis} side.

\subsection{Characterization of dynamics via ejection time}
\label{sec:res:pore}

\begin{figure}
\centering
\includegraphics[width=0.49\linewidth]{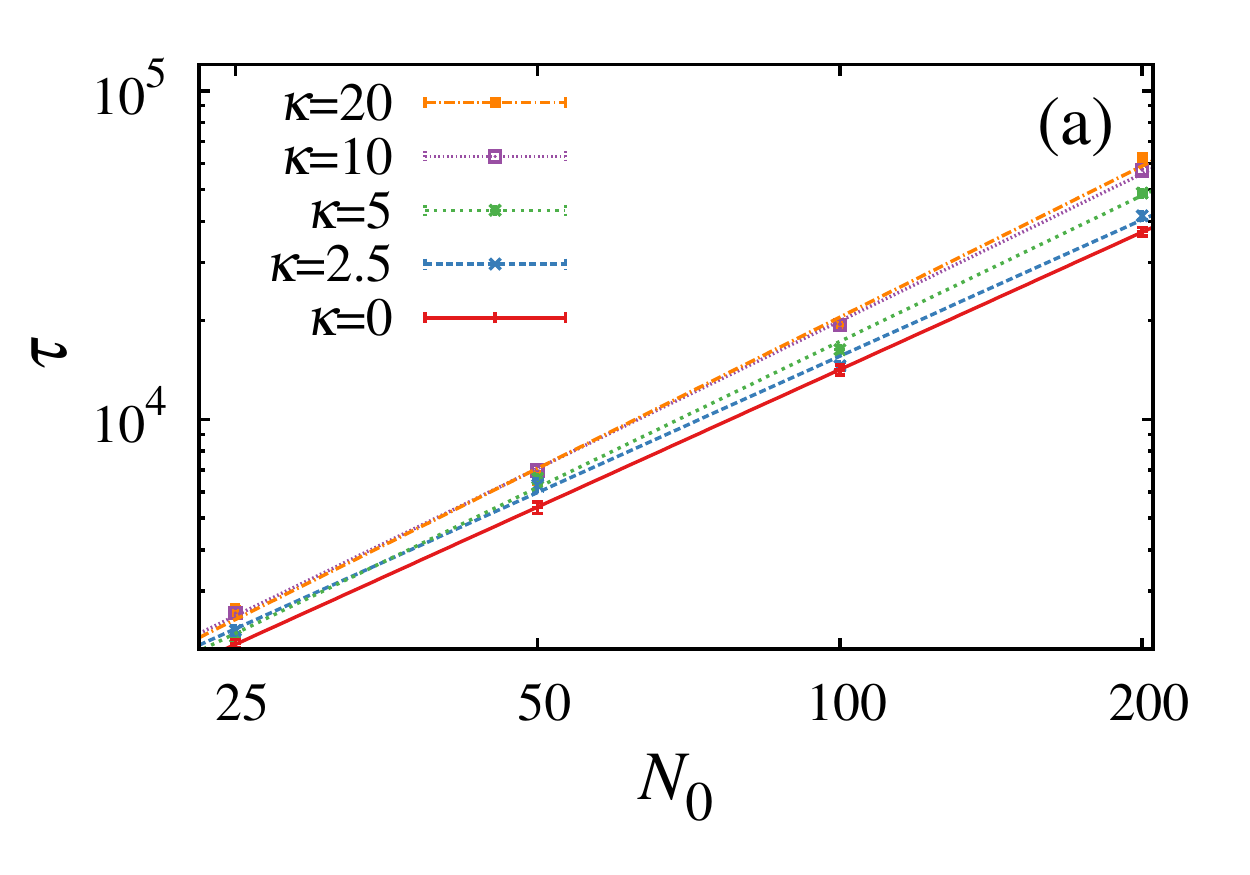}
\includegraphics[width=0.49\linewidth]{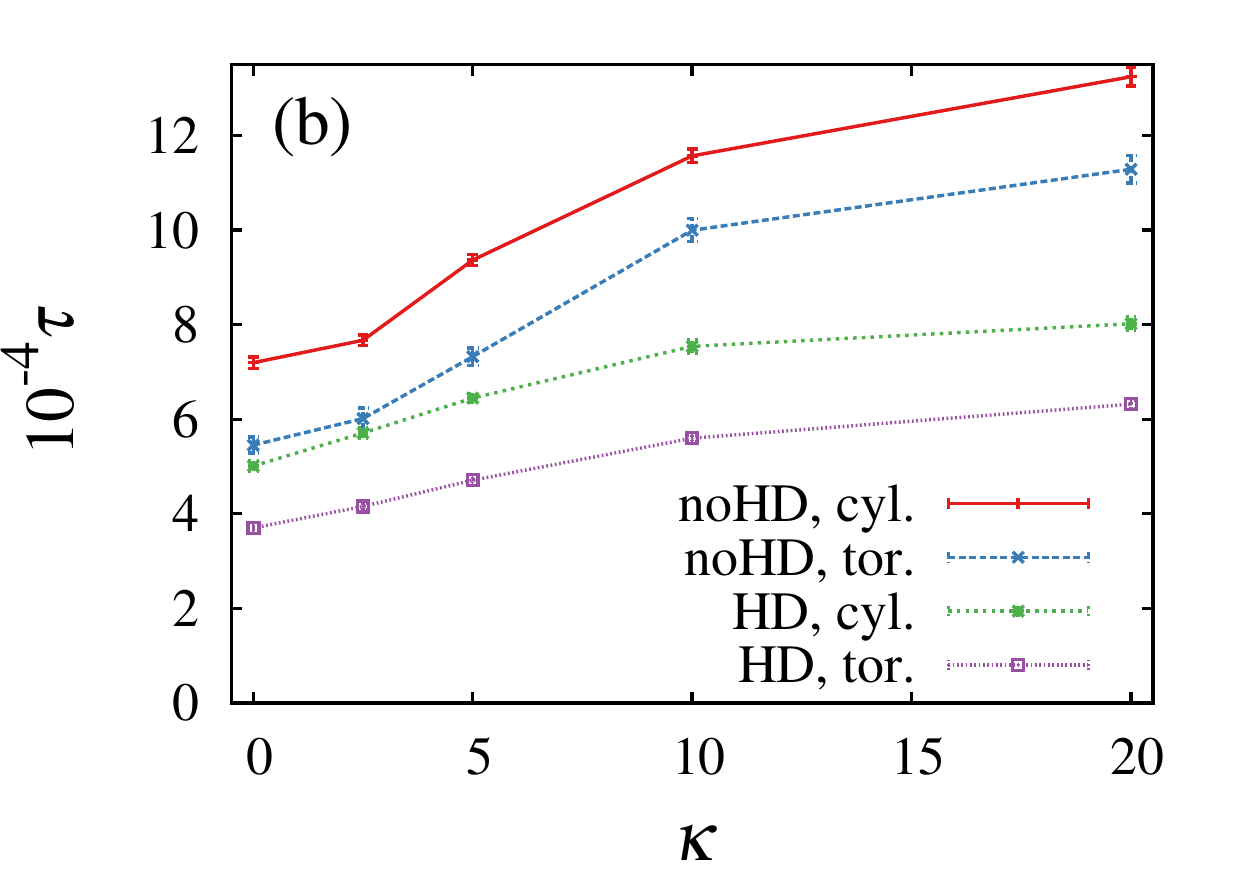}
\caption{(Color online) (a) Torus pore. Hydrodynamics enabled. Ejection time $\tau$ vs polymer length $N_0$ for different \ka and for $N_0 = 25$, $50$, $100$, and $200$. Lines show the relation $\tau\sim N_0^\beta$ fitted to the data. Logarithmic scale. The standard deviations of the mean are of the same order as the size of the points. (Values of $\beta$ are shown in Table~\ref{tab:tauVsN}.) (b) Ejection time $\tau$ as a function of \ka for the cylinder and torus pores with and without hydrodynamics for $N_0=200$.}
\label{fig:tauVsN}
\end{figure}

\begin{table}
\caption{The fitted exponents in the relation $\tau\sim N_0^\beta$ for different values of bending parameter \ka with and without hydrodynamics (HD). The torus and the cylinder pore.}
\label{tab:tauVsN}
 \begin{tabularx}{\linewidth}{X|XX|XX|}
\hline\hline
& \multicolumn{2}{c|}{torus} & \multicolumn{2}{c|}{cylinder}\\
\ka & noHD  & HD  & noHD  & HD  \\\hline
0   & 1.43  & 1.39 & 1.46 & 1.35 \\
2.5 & 1.48  & 1.38 & 1.41 & 1.37 \\
5   & 1.53  & 1.48 & 1.43 & 1.39 \\
10  & 1.57 & 1.49 & 1.45 & 1.40 \\
20  &  1.48 & 1.53 & 1.35 & 1.39 \\
\hline\hline
\end{tabularx}
\end{table}

We mainly use our implementation of the low-friction torus pore in order to more sensitively capture the dynamical characteristics of ejection. To validate the pore model and further our understanding of the ejection dynamics we make some comparison to the higher-friction cylinder pore used in our previous work.

Figure~\ref{fig:tauVsN}~(a) shows the ejection time $\tau$ as a function of polymer length $N_0$ for various values of \ka for the torus pore with included hydrodynamics. $\tau$ is seen to increase monotonically with \ka for all $N_0$. Hence, for polymers of increased rigidity the effects of increased friction and less malleable conformations dominate over the increased force due to confinement driving the polymer out of the capsid.

For all \ka,  ejection times follow closely the scaling relation $\tau\sim N_0^\beta$. Scaling of $\tau$ with $N_0$ for the cylinder pore is equally good (not shown). $\tau$ as a function of rigidity $\kappa$ are shown in Fig.~\ref{fig:tauVsN}~(b). The obtained $\beta$ are given in Table~\ref{tab:tauVsN}. Due to volume correction, Eq.~\eqref{eq:density}, $\beta$ for $\kappa = 0$  are slightly greater than those reported in~\cite{piili_capsid2}. The two  pores give consistent results. For the cylinder pore the $\beta$ exponents are smaller due to the larger pore friction, increasing of which takes $\beta$ closer to $1$. $\beta$ is seen to increase with $\kappa$ for $\kappa \le 10$. This trend breaks down for $\kappa = 20$ indicating a change in the dynamics.

Hydrodynamics is seen to speed up equally the ejection of polymers for all \ka. Both the hydrodynamic interactions and larger \ka increase the polymer's correlation length but have opposite effect on ejection rate. The correlation length increases by hydrodynamics due to momentum mediated by the solvent, which assists polymer motion. In contrast, the increased correlation length due to bending rigidity does not involve momentum transfer but only increases friction exerted on the moving polymer segment. In the presence of hydrodynamics, where correlation length is already increased, increasing \ka does not increase correlation length further as in the absence of hydrodynamics. Consequently, the ejection time does not increase with $\kappa$ as much when hydrodynamics is included, as seen in Fig.~\ref{fig:tauVsN}~(b). Hydrodynamics is also seen to reduce $\beta$, the only exception being the case $\kappa = 20$, which was found to differentiate from ejections for smaller $\kappa$ also in how $\beta$ changes with $\kappa$. The reason for this difference can be understood by studying the cumulative waiting times, see Section~\ref{sec:scaling}.

\subsection{Pore force}
\label{sec:poreforce}

\begin{figure}
\def \wid {0.49}
\includegraphics[width=\wid\linewidth]{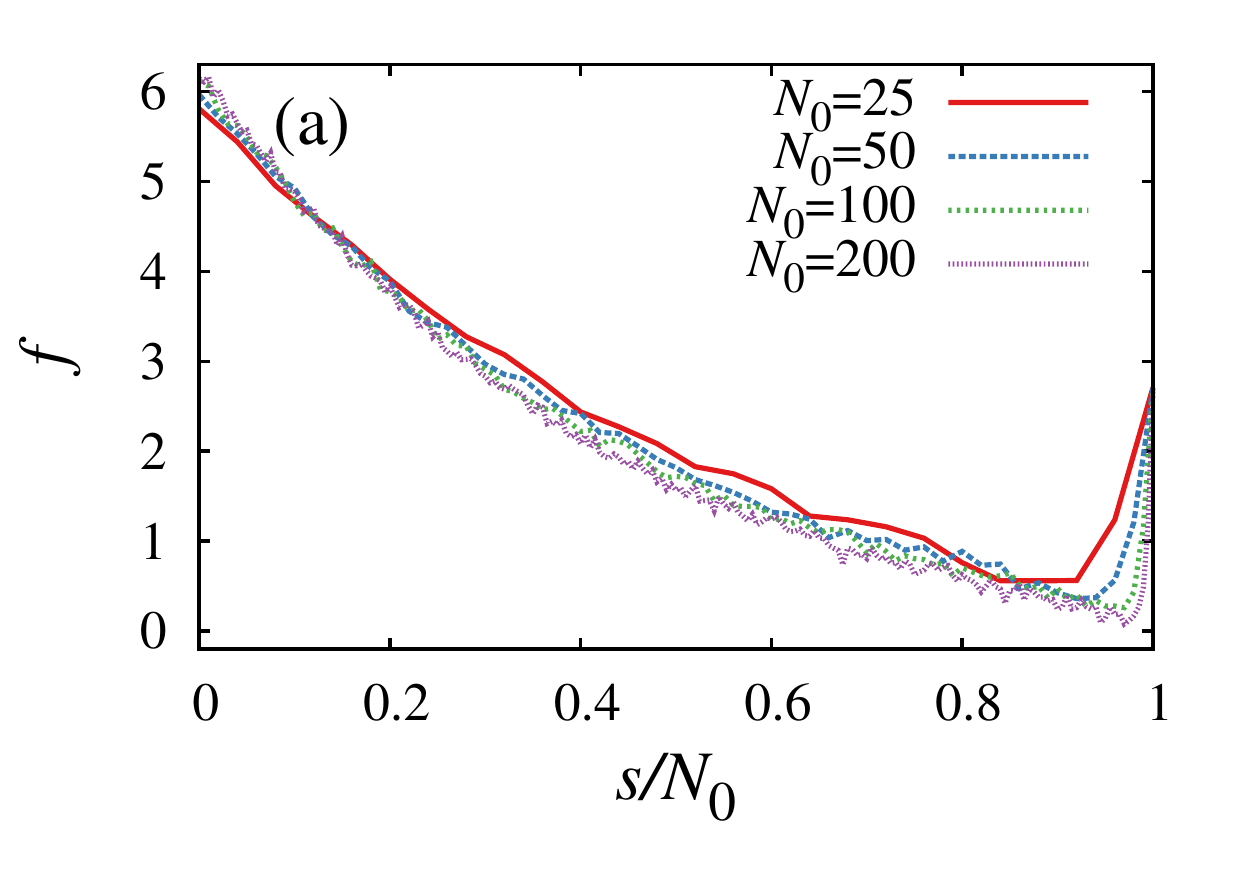}
\includegraphics[width=\wid\linewidth]{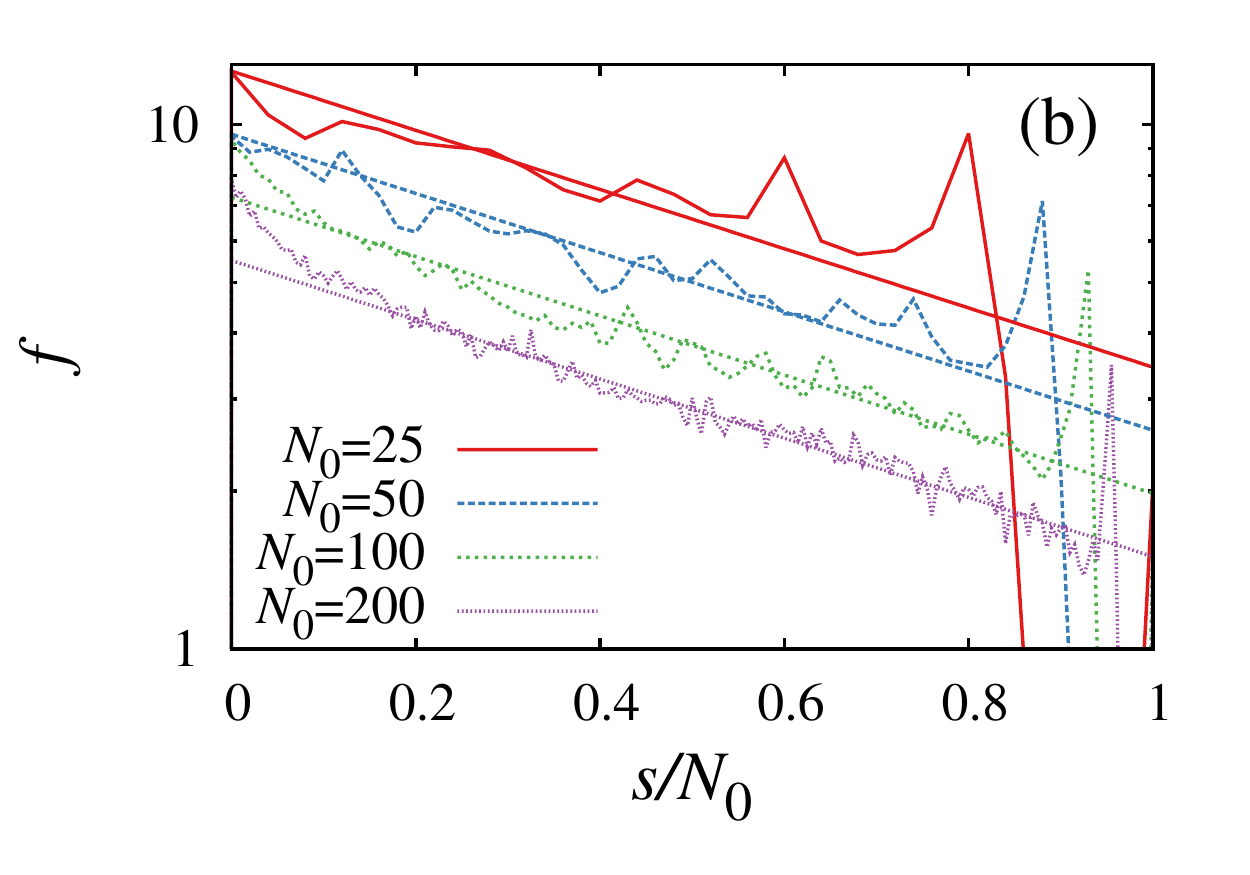}
\caption{(Color online) Equilibrium force $f$ measured at the pore as a function of normalized number of ejected monomers $s/N_0$ for (a) $\kappa=0$ and (b) $\kappa = 20$. the lines show  $f = 12.4883 \cdot (1/N_0)^{0.4}\exp{\left(-1.3s/N_0\right)}$.}
\label{fig:f_kappa}
\end{figure}

Figure~\ref{fig:f_kappa}~(a) shows the equilibrium force $f$ measured at the pore for fully flexible ($\kappa=0$) polymers as a function of the number of ejected monomers normalized by the polymer length $s/N_0$. $f$ for different $N_0$ essentially collapse when plotted this way. This means that the pore force is essentially a function of density inside the capsid. In addition, $f$ is  closely reminiscent of the inverse of the resulting waiting time profile $1/t_w$ where $t_w(s)$ is the elapsed time between the last ejections of monomers $s$ and $s+1$~\cite{piili_capsid,piili_capsid2}. Consequently, the cumulative waiting time $t(s) = \int_0^s t_w(s')ds'$ in Fig.~\ref{fig:tVsN} is closely related to $f(s)$ for $\kappa = 0$.  

As \ka increases $f$ is no longer constant for constant $s/N_0$. With $\kappa=20$, $f$ for different $N_0$ has roughly the dependence $f \propto (1/N_0)^{0.4} \exp(-1.3s/N_0)$, see Fig.~\ref{fig:f_kappa}~(b). Due to starting all ejections from a constant monomer density, $N_0$ is proportional to $V$, where $V$ is the capsid volume, see Eq.~(\ref{eq:density}). Hence, $f \propto (1/V)^{0.4}$. For the constant $\kappa = 20$ the persistence length is constant for all $N_0$. The force exerted on the capsid wall increases with $\lambda_p/R_0$. Writing $f \propto (\lambda_p/R_0)^\gamma$  and using $f \propto 1/V^{0.4}$ gives $\gamma \approx 1.2$. A value $\gamma \gtrsim 1$ is plausible. For one semiflexible beam crossing through the midpoint of the capsid and touching opposite walls one would expect $\gamma$ of the order of $1$. In the actual polymer conformation with intertwined semiflexible lobes pressing against each other and the walls the increase of $f$ with $\lambda_p$ is expected to be superlinear. The dependence $f \propto (1/N_0)^{0.4}$ implies the breakdown of the above-described direct relation of $f$ and $1/t_w$ for more rigid polymers. In other words, as $\kappa$ increases ejection dynamics is less directly related to the force measured at the pore, as already evidenced by the rigid polymers ejecting slower than more flexible polymers in spite of the spring force increasing with rigidity, see Fig.~\ref{fig:tauVsN}.

\subsection{Cumulative waiting times: cross-over to scaling}
\label{sec:scaling}

\begin{figure}
\includegraphics[width=0.49\linewidth]{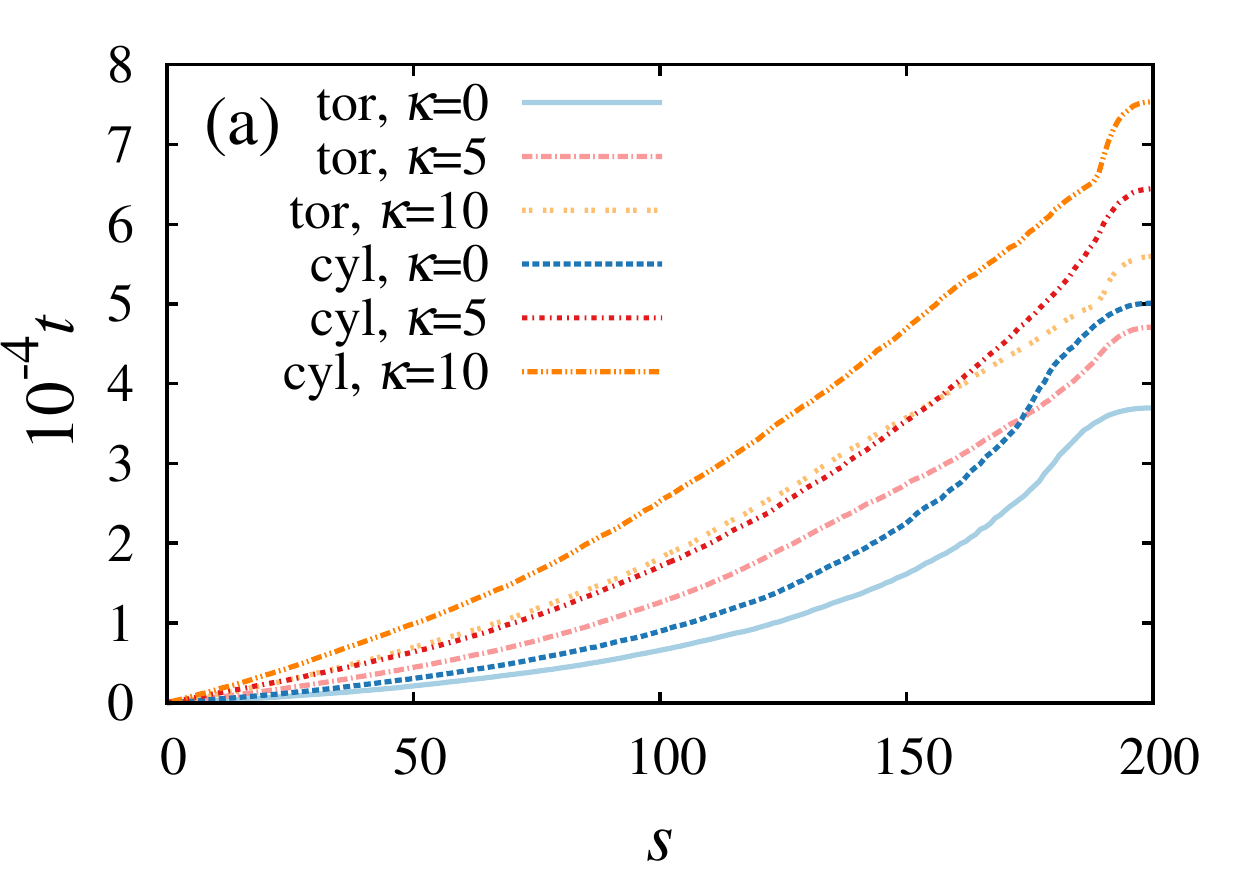}
\includegraphics[width=0.49\linewidth]{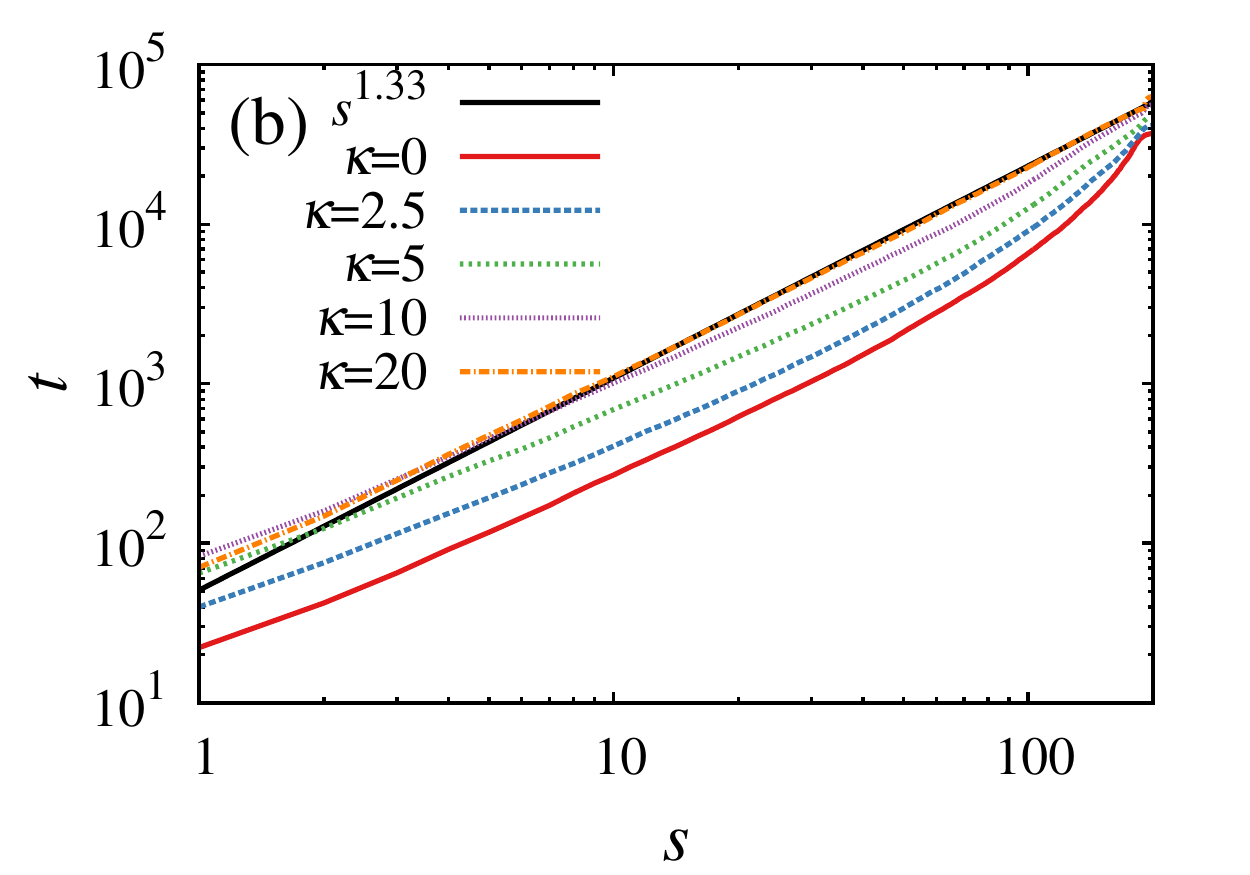}
\caption{(Color online) (a) Cumulative waiting times $t(s)$ for the torus (tor.) and cylinder (cyl.) pore models using $\kappa = 0$, $5$, and $10$. $N_0=200$, $\rho_0=0.75$. (b) Torus pore. Cumulative waiting times for polymers of length $N_0=200$ and for different \ka. The black solid line shows the scaling $t\sim s^{1.33}$.}
\label{fig:cylinderVsTorus}
\end{figure}

\begin{figure*}
\def \wid {0.49}
\includegraphics[width=\wid\linewidth]{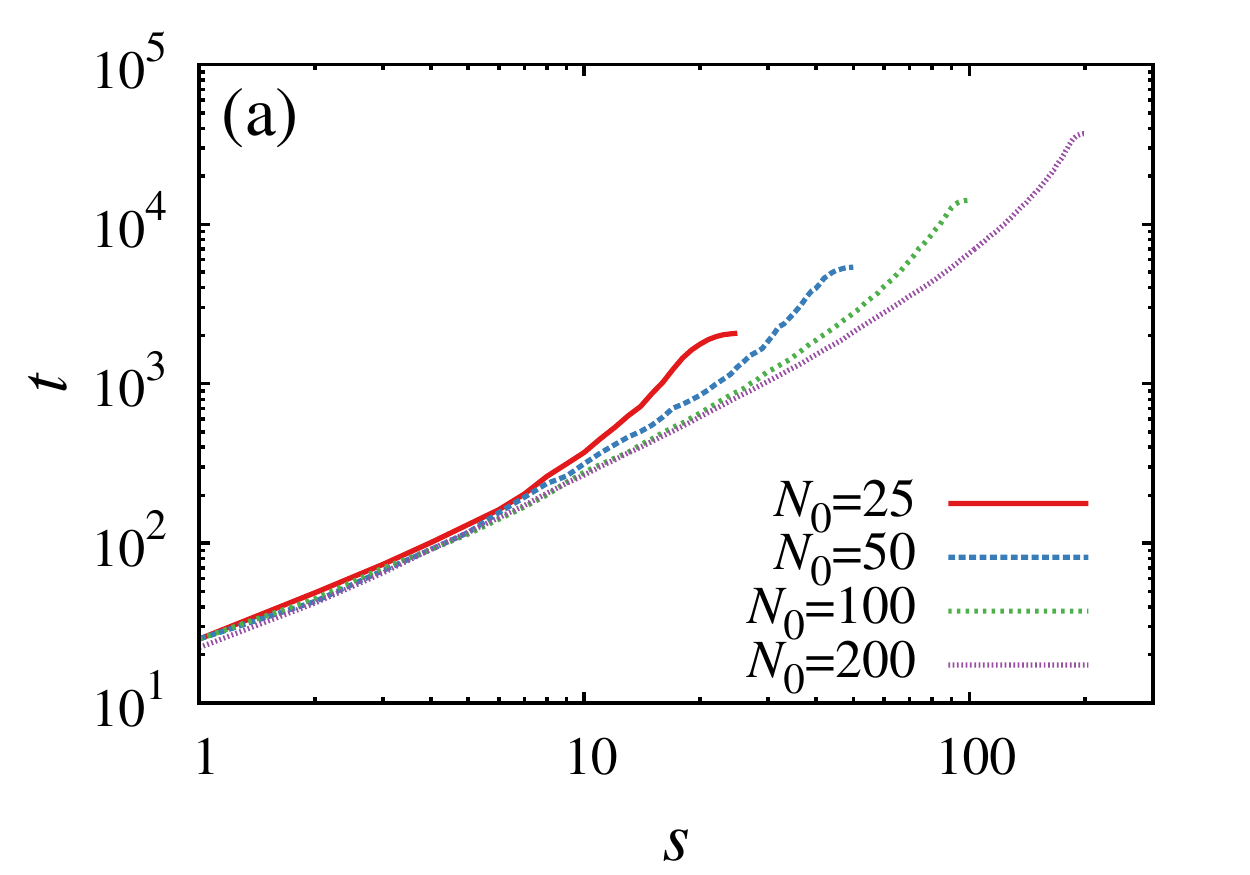}
\includegraphics[width=\wid\linewidth]{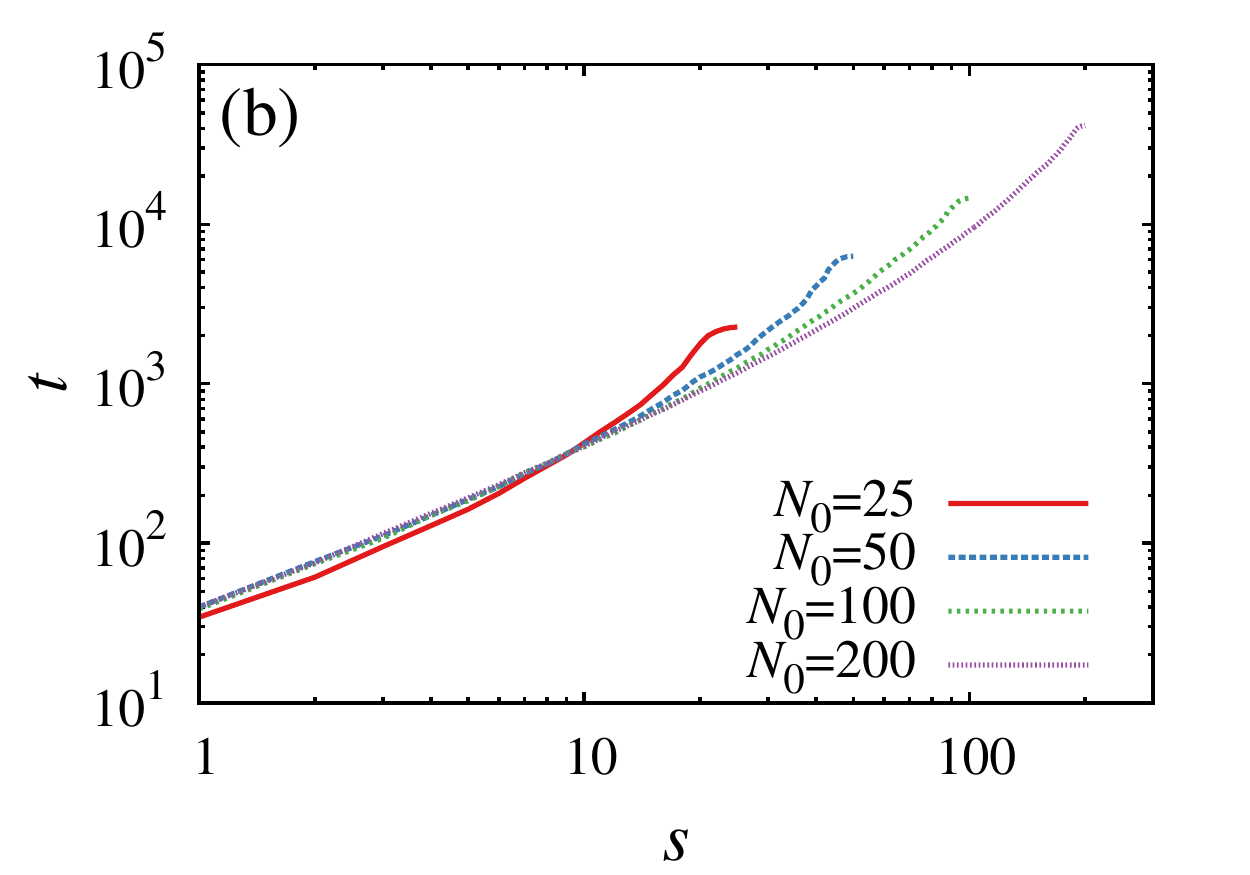}
\includegraphics[width=\wid\linewidth]{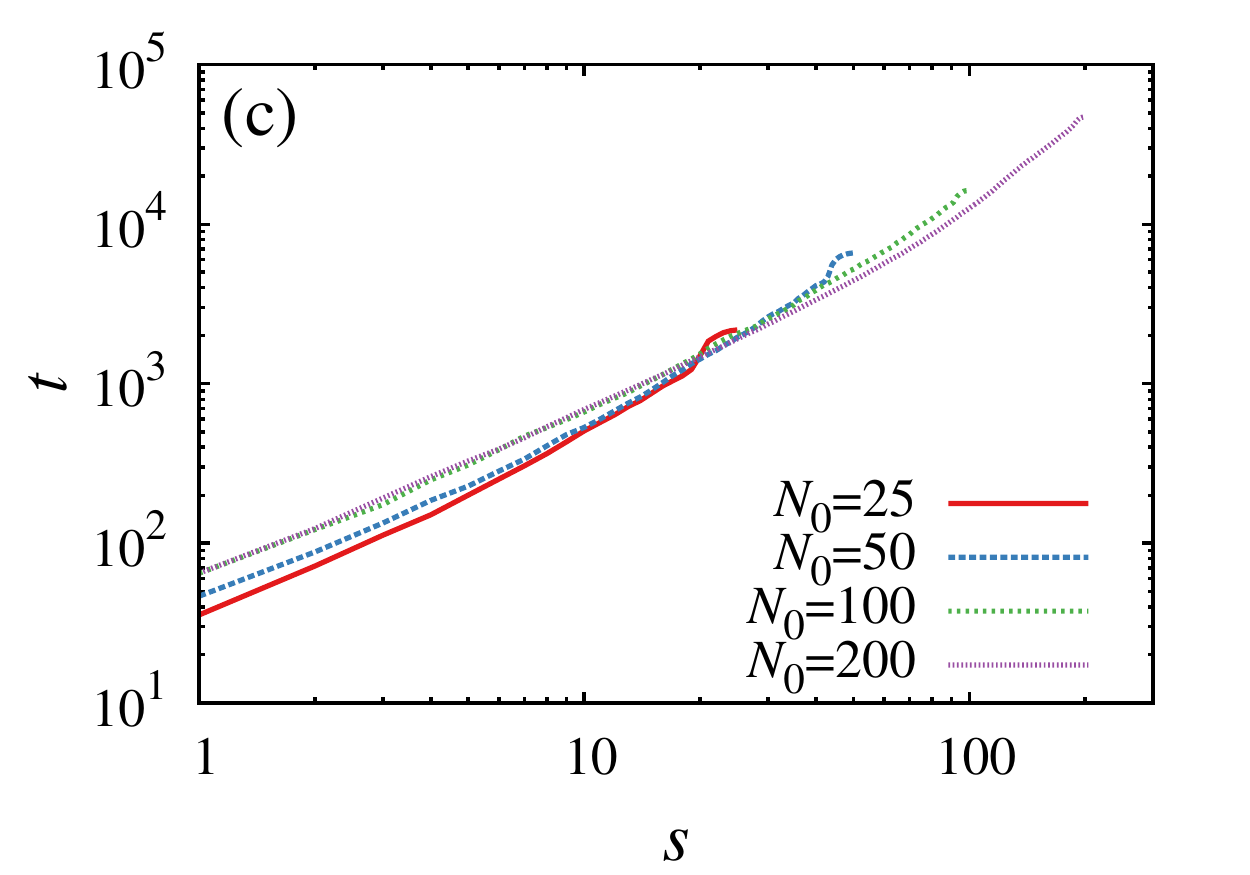}
\includegraphics[width=\wid\linewidth]{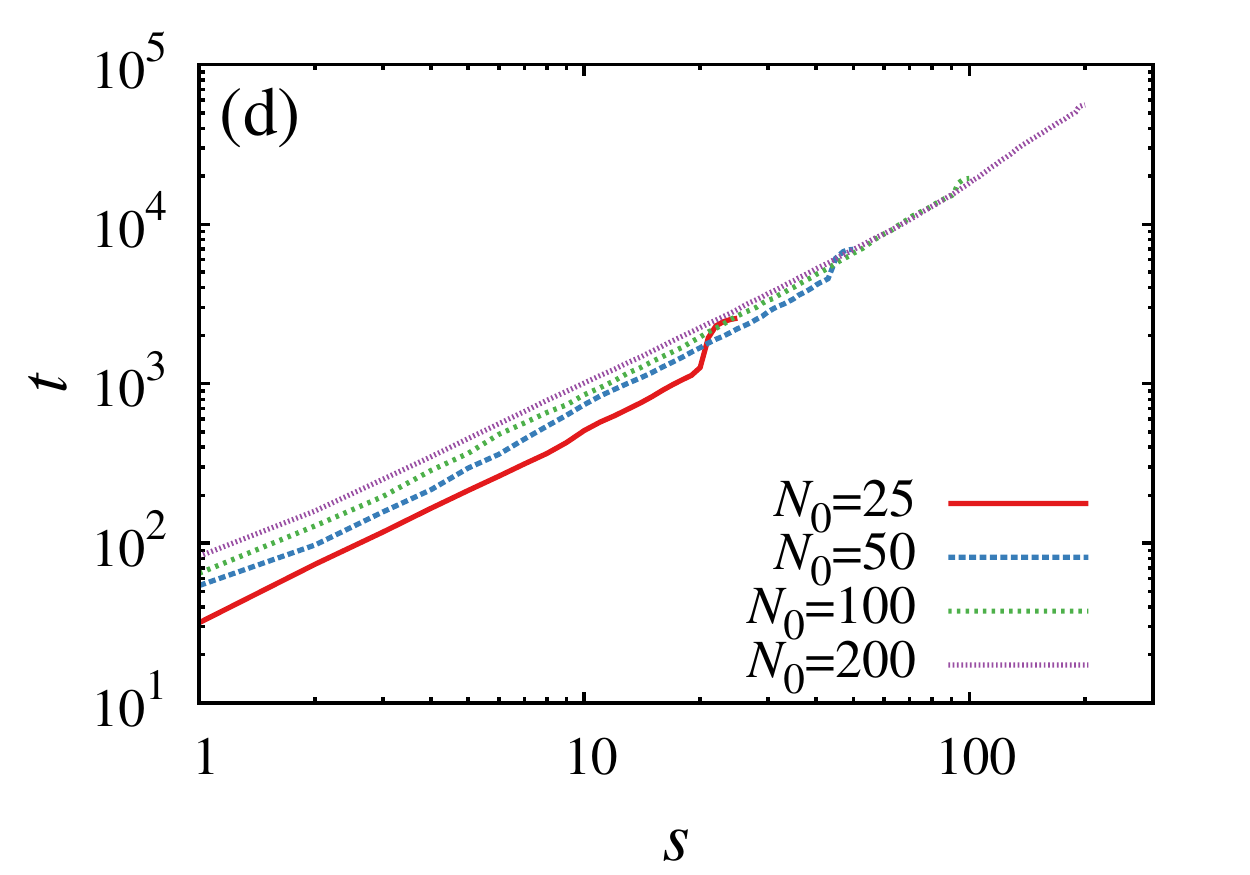}
\includegraphics[width=\wid\linewidth]{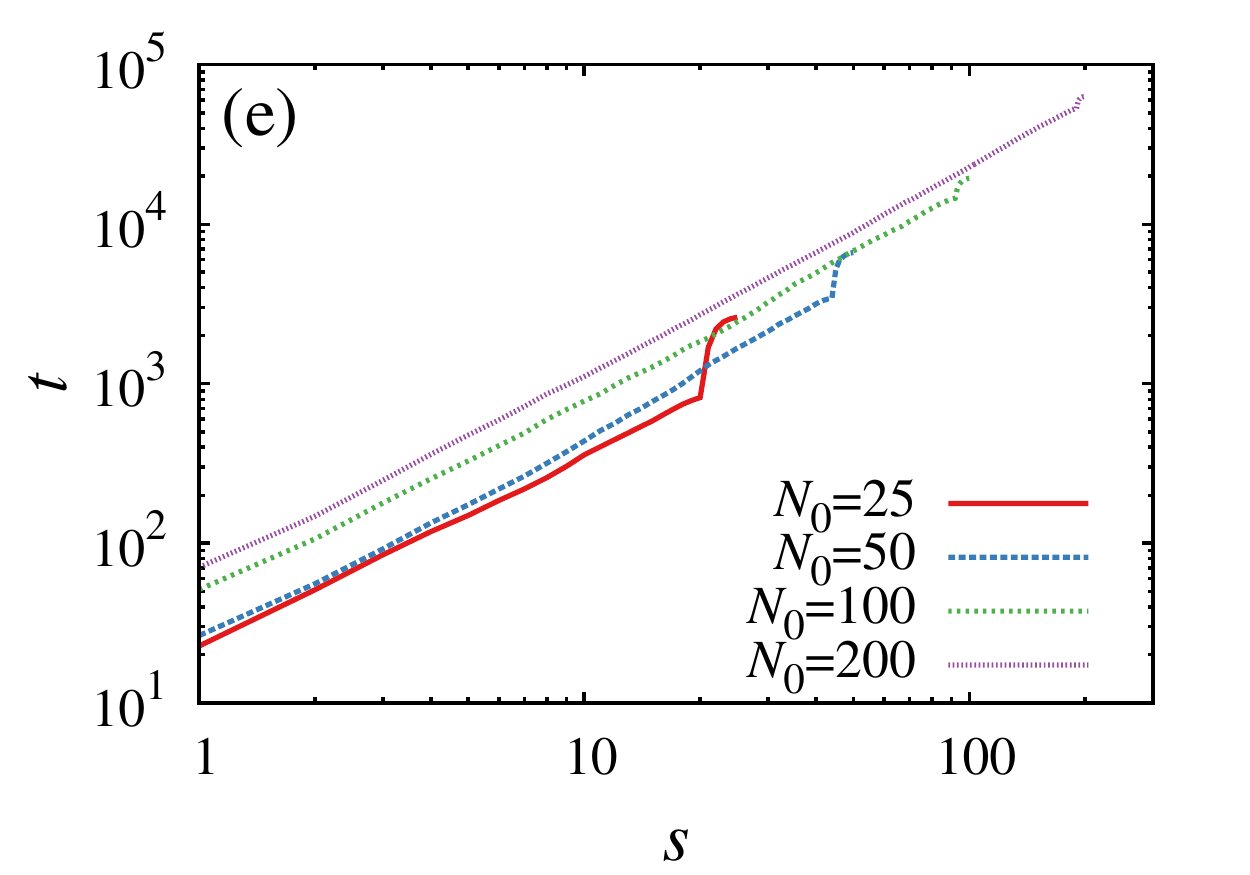}
\includegraphics[width=\wid\linewidth]{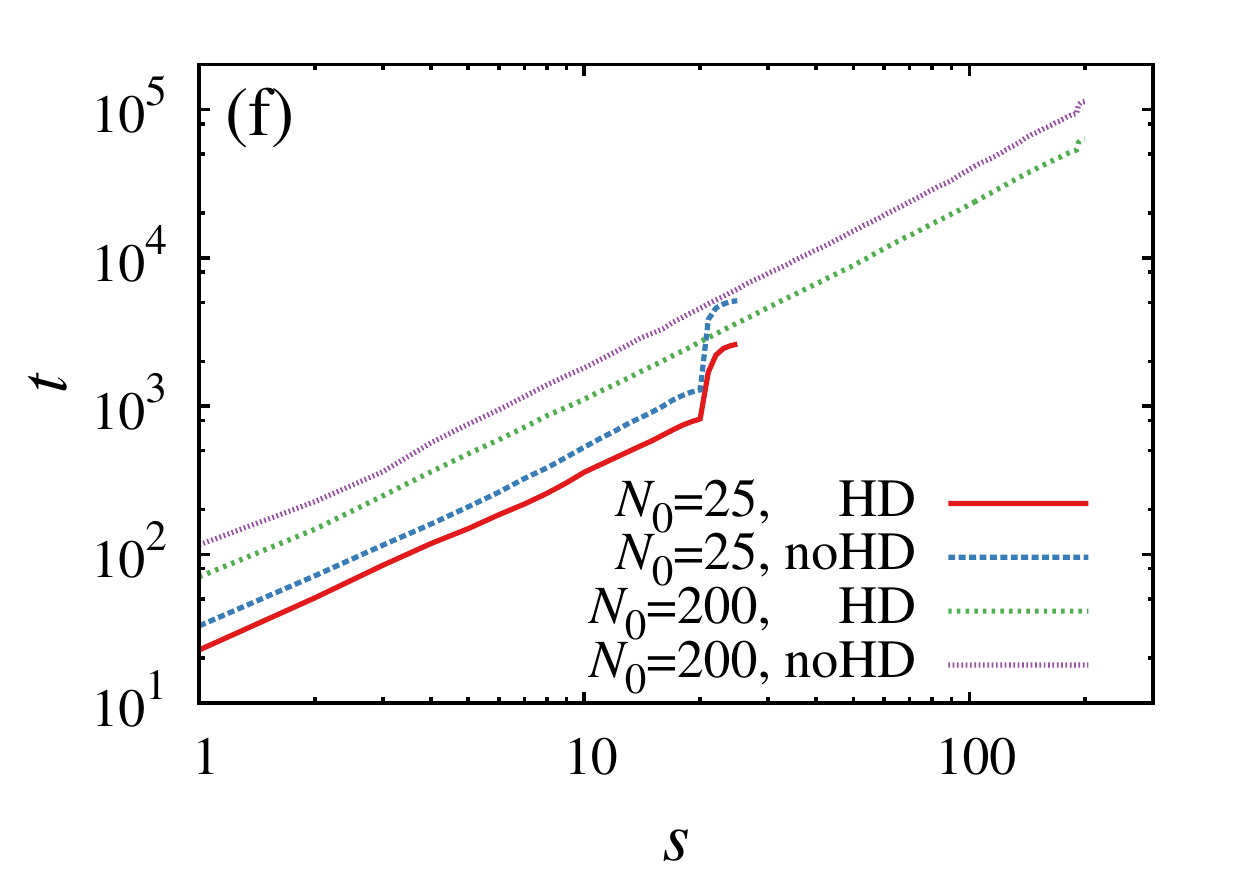}
\caption{(Color online) Cumulative waiting times $t$ vs $s$ in logarithmic scale. Torus pore. (a) - (e) Hydrodynamics included, $\kappa=0,2.5,5,10,20$. (f) With and without hydrodynamics, $\kappa = 20$.}
\label{fig:tVsN}
\end{figure*}

We define the cumulative waiting time $t(s)$ as the time it takes for the bead of index $s$ to eject the capsid for the last time. In spite of the apparent scaling of $\tau$ with $N_0$, $t$ does not scale with $s$ for fully flexible polymers, as we have shown previously~\cite{piili_capsid, piili_capsid2}. For verification of the pore models, Fig.~\ref{fig:cylinderVsTorus}~(a) shows $t(s)$ obtained using the two different pores for different \ka is shown for polymers of length $N_0=200$. The forms of $t(s)$ using the two pores are seen to be quite similar.

It is seen in Fig.~\ref{fig:cylinderVsTorus}~(b) that similarly to the fully flexible polymers, $t$ does not scale with $s$ for moderate and medium \ka. However, for sufficiently rigid polymers scaling emerges. For $\kappa = 20$ perfect scaling $t \sim s^{1.33}$ is obtained for all $N_0$. Only at the end of the ejection there is the final diffusion of the polymers out of the capsid, during which $t(s)$ differs from the previous scaling relation. It is also evident from Fig.~\ref{fig:cylinderVsTorus}~(a) that $t(s)$ increase with $\kappa$.

Fig.~\ref{fig:tVsN} shows cumulative waiting times for each studied polymer length such that \ka grows from figure (a) to (e) from $\kappa=0$ to $\kappa=20$. When $\kappa=20$ for the main part of the ejection where the monomer density inside the capsid is sufficiently high that the polymer is being pushed out of the capsid, the cumulative waiting times scale like $t \sim s^{1.33}$, whether hydrodynamics is included or not. Regarding the apparent scaling of the total ejection time $\tau \sim N_0^\beta$, the increase of $\beta$ with $\kappa$ for $\kappa \le 10$ no longer holds for $\kappa = 20$, since transition to a different dynamical regime has taken place, see Table~\ref{tab:tauVsN}.

Inclusion of hydrodynamics was seen to reduce $\beta$ for all but the polymers of largest rigidity $\kappa = 20$, see Table~\ref{tab:tauVsN}. This effect comes from the final diffusive part of the ejection. For $\kappa=20$ the diffusion at the final stage takes clearly longer for short polymers in the absence of hydrodynamics, see Fig.~\ref{fig:tVsN}~(f). The friction, which increases with $\kappa$, slows down the final diffusion of the polymer from the capsid. Hydrodynamic interactions reduce the friction of the diffusing rigid polymer tail. As seen in Fig.~\ref{fig:tVsN}~(f), for $\kappa = 20$ the contribution of the final diffusion on the total ejection time is largest for the shortest polymers, which reduces $\beta$. Since the friction of the diffusing tail is larger in the absence of hydrodynamics, $\beta$ is larger when hydrodynamics is included.

For $\kappa \le 10$ the inclusion of hydrodynamics, which reduces friction outside the pore, leads to enhanced relative friction due to the pore. Hence, the smaller $\beta$ when including hydrodynamics and when using the cylinder pore comes from the same origin. This is similar as in translocation where increasing the pore friction trivially takes the scaling toward linear~\cite{lehtola_epl}.

As \ka increases, there is a transition from exponential increase of $t$ with $s$ \cite{piili_capsid,piili_capsid2} to a regime where perfect scaling of $t \sim s^{1.33}$ emerges for the main part of the ejection. This transition has the characteristics of a dynamical phase transition, where \ka acts as a control parameter. The scaling exponent equals $1 + \nu$, where $\nu = 1/3$ would be expected for a polymer driven through a pore from a spherical conformation in a way that relaxation does not appreciably change the conformation and the conformation is homogenous. 

Since simulated ejections for different $N_0$ start from a constant monomer density, Eq.~(\ref{eq:density}), the capsid radius scales as $R_0 \sim N_0^{1/3}$. A semi-rigid polymer densely packaged inside a capsid is in a conformation where intertwined lobes fill the spherical volume. The angles and directions of these lobes vary randomly. Although for large $\kappa$ ejection is slowed down, the conformations do not appreciably change due to relaxation for $\kappa = 20$, since the persistence length is clearly greater than the capsid diameter. (From Table~\ref{tab:kappaVsPersistence}, $\lambda_p \approx 21.9$ for $\kappa = 20$ and $N_0 = 200$. The diameter of the capsid used for this polymer is roughly $8$.) Hence, the measured $t(s)$, which is an average over ejections, reflects the initial random conformations. The spherical scaling of the ejection of the randomly oriented rigid segments would seem to result from the spherical confinement.

\subsection{Scaling functions}
\label{sec:sc_functions}

\begin{figure*}
\def \wid {0.49}
\includegraphics[width=\wid\linewidth]{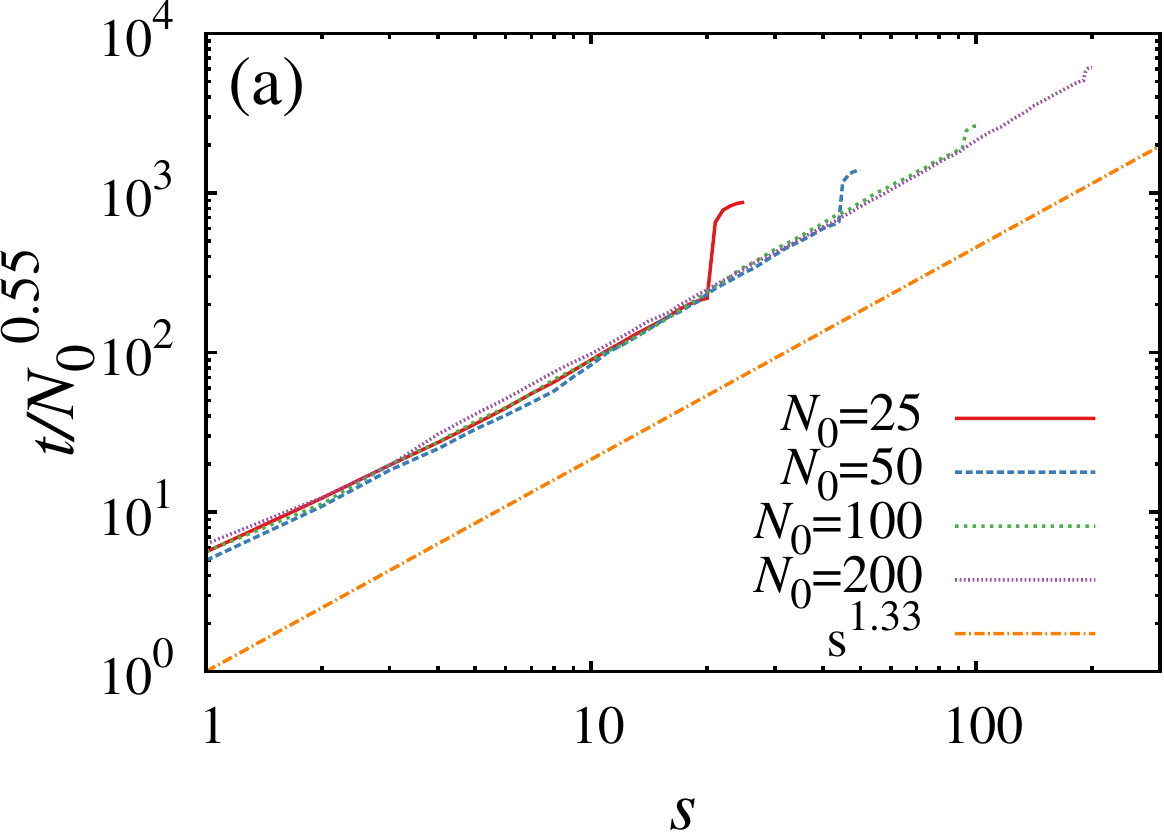}
\includegraphics[width=\wid\linewidth]{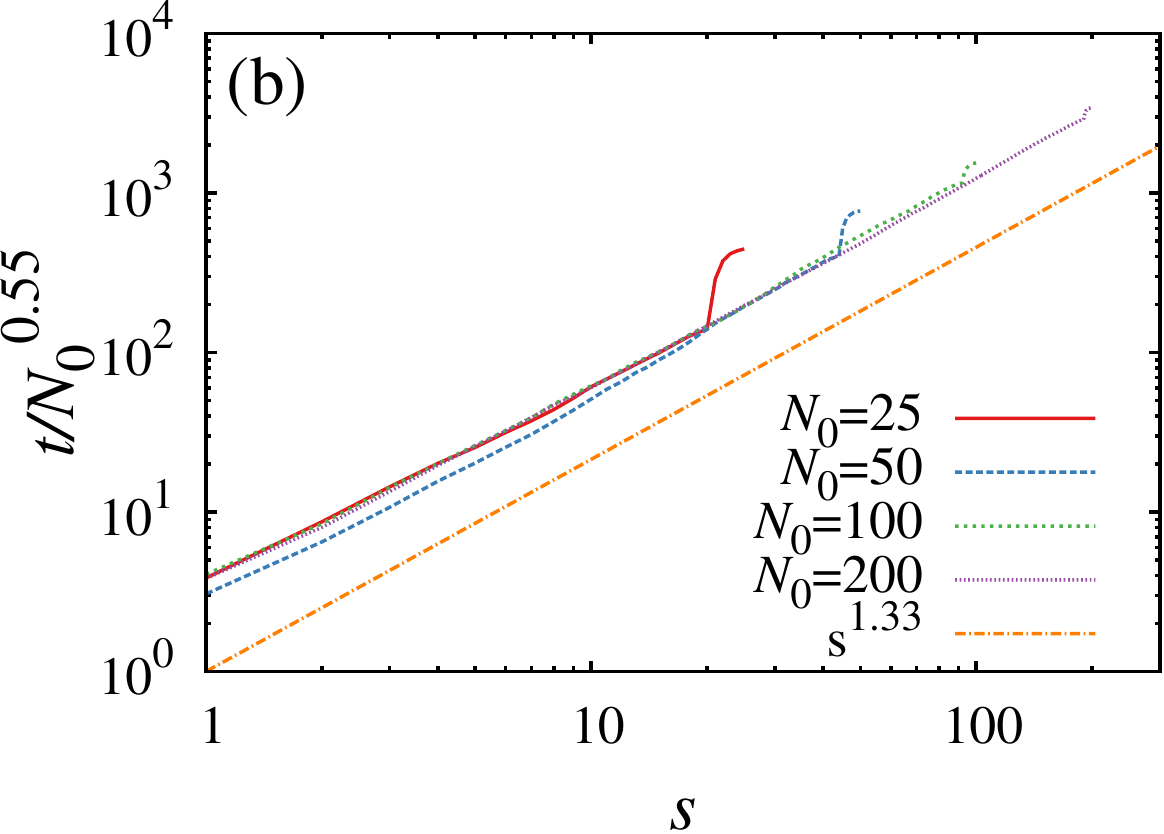}
\caption{(Color online) Scaled cumulative waiting times $t(s)/N_0^{0.55}$ for $N_0 = 25$, $50$, $100$, and $200$ with $\kappa=20$. (a) No hydrodynamics. (b) Hydrodynamics included. The separate dashed lines shows the scaling $\sim s^{1/3}$.}
\label{fig:t_collapse}
\end{figure*}

\begin{figure*}
\def \wid {0.49}
\includegraphics[width=\wid\linewidth]{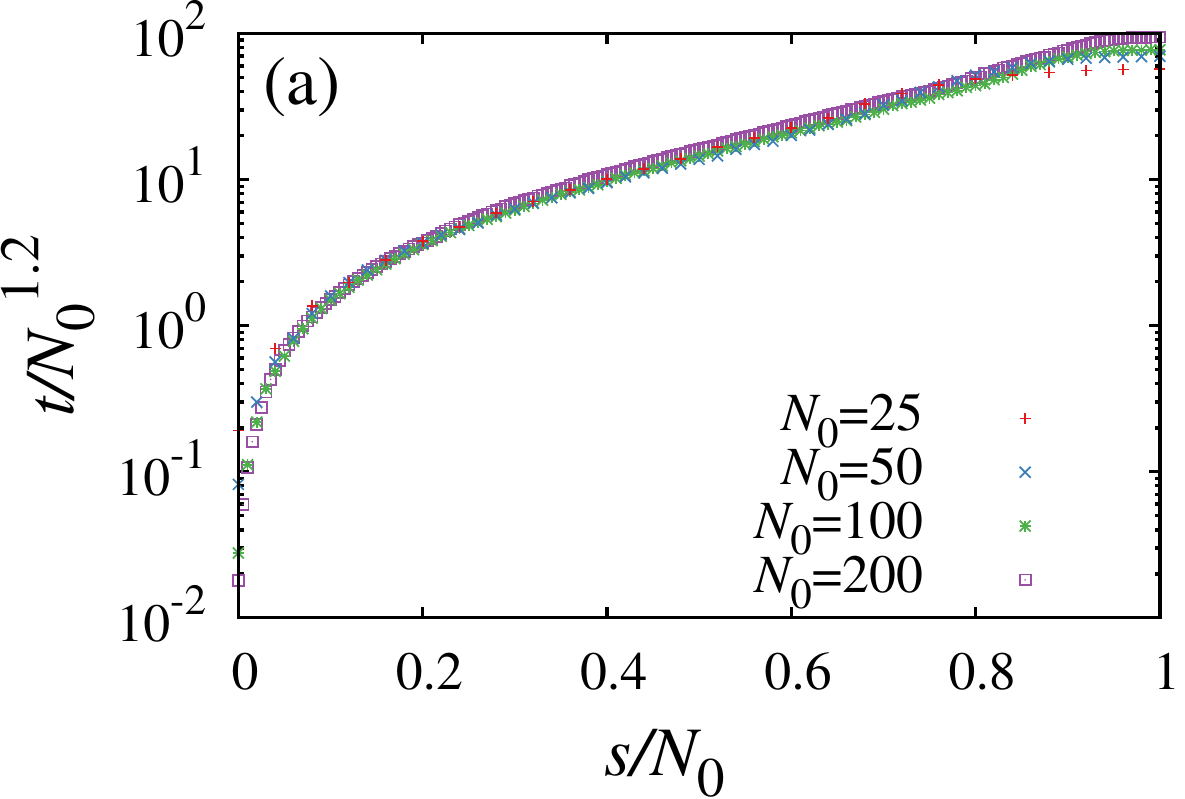}
\includegraphics[width=\wid\linewidth]{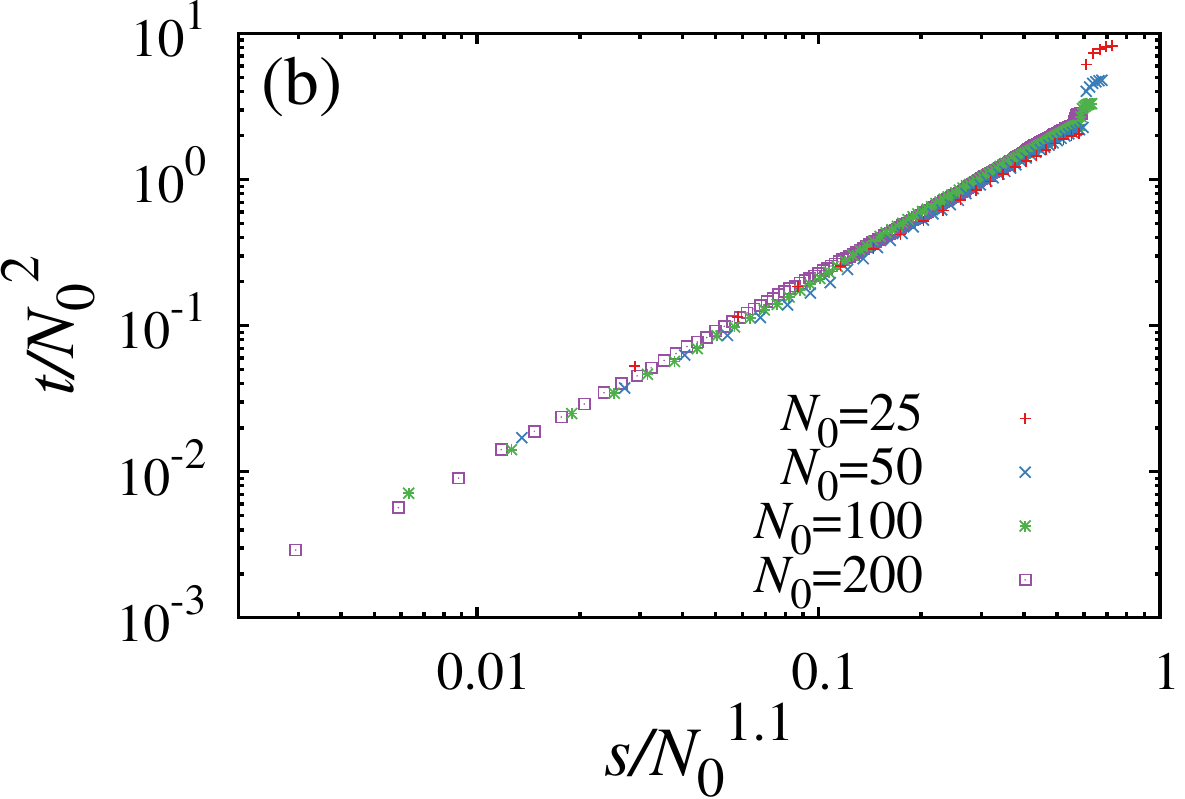}
\caption{(Color online) Data collapse for (a) $\kappa = 0$ (b) $\kappa = 20$.}
\label{fig:data_collapse}
\end{figure*}

For $\kappa=20$, the force $f$ multiplied by $N_0^{0.4}$ plotted as a function of $s/N_0$ approximately collapses in a large range, see Fig~\ref{fig:f_kappa}~(b). Therefore, it is reasonable to expect that such a data collapse exists also for the cumulative waiting time $t$. This, indeed, is the case. Fig.~\ref{fig:t_collapse} shows the cumulative waiting times $t(s)$ for different $N_0$ scaled by $1/N_0^{0.55}$. The parts of $t(s)/N_0^{0.55}$ scaling as $\sim s^{1.33}$ are seen to fall on top of each other. The scaling of force $f \sim 1/N_0^{0.4}$ for each $s/N_0$ naturally shows as scaling in the cumulative waiting time magnitudes. The cumulative waiting time for sufficiently rigid polymers in the regime where the polymer is driven by the pressure inside the capsid can thus be written as 
\begin{equation}
t(s) \propto N_0^{0.55} s^{1.33} 
\end{equation}
both in the absence and presence of hydrodynamics. The prefactor $N_0^{0.55}$ indicates that it may be possible to select capsid volumes for different $N_0$ such that the scaling parts of $t(s)$ have the same magnitude. The initial condition, which for fully flexible polymers was judiciously chosen as  starting from constant monomer density, needs to be modified to take into account of the relation of the persistence length to the capsid radius and the polymer length, among other things.

Finally, we note that for each $\kappa$ a single scaling function can be found to describe the measured cumulative waiting times of polymers of different $N_0$. Fig.~\ref{fig:data_collapse} shows the data collapse for the minimum and maximum rigidities used in our studies. The cumulative waiting times can be expressed by the corresponding scaling functions as $t(s,N_0) = N_0^{1.2}\Gamma(s/N_0)$ for $\kappa = 0$ and $t(s,N_0) = N_0^2\Psi(s/N_0^{1.1})$ for $\kappa = 20$. It should be borne in mind that precise exponent values can be obtained only when finite-size effects, such as the density correction, are taken properly into account. A scaling function that would describe $t$ in a small range of $\kappa$ around $\kappa = \kappa_c$, where $\kappa_c$ is the value at which the transition to $t \sim s^{1.33}$ occurs, could serve as a starting point for a renormalization group approach for a universal description of the dynamical transition as a function of polymer rigidity.

\section{Conclusion}\label{sec:con}

We have studied the ejection of semi-flexible polymers from a spherical capsid and under initial strong confinement. The ejection simulations were performed by stochastic rotation dynamics, which is  molecular dynamics based method where inclusion of hydrodynamic modes is possible. The capsid was implemented via simple boundary conditions and the effect of geometrical pore friction was minimized using a torus pore with perfectly round edges. This pore implementation was validated by comparison to the previously used cylinder pore. The initial conformations were generated by capsid injection using Langevin dynamics.

We found that the rate at which polymers are packaged inside the capsid affects the ejection time. The conformations that were allowed less time for equilibration during packaging ejected faster. This characteristics is likely related to the deeper local energy minimima of the conformations that were allowed to relax for a longer time. The time allowed for equilibration was chosen such that increasing this time further did not change the ejection time appreciably. By measurements of the radius of gyration for the polymer segment outside the capsid during ejection we found that all contribution to the change in the dynamics with the polymer length $N_0$ comes from inside the capsid. 

Unlike in the case of fully flexible polymers, we found that for semiflexible polymers with $\kappa=20$ the force $f$ measured at the pore has no clear relation to the measured monomer ejection times. $f$ showed clear dependence on the polymer length such that the force magnitudes as a function of $s/N_0$ scaled like $f \sim N_0^{-0.44}$.

There are basically two regimes in ejection dynamics. For the main part monomer density in the capsid is sufficiently high that the polymer is being driven out through the pore. In the second regime at the end of ejection the remaining segment diffuses out of the capsid. For low $\kappa$ the regimes are overlapping. This overlap diminishes as $\kappa$ is increased and scaling of the cumulative waiting time $t$ with $s$ emerges. For $\kappa = 20$ the regimes are clearly separate and perfect scaling of $t(s) \sim s^{1.33}$ is obtained for the driven regime both in the absence and presence of hydrodynamic interactions. The scaling exponent is precisely $1 + \nu$, where $\nu = 1/3$ is the scaling for the radius $R$ of spherically confined polymer conformations $R \sim N_0^\nu$. We gave some arguments for obtaining $t(s) \sim s^{1+\nu}$ when averaged over many individual ejections of sufficiently rigid polymers.

Hydrodynamics was found to speed up the ejection at all stages. The speed-up was strongest in the diffusion regime at the end of the process, where hydrodynamic interactions lower the friction caused by the rigid tail inside the capsid. We found a perfect collapse for the scaled cumulative waiting times in the driven regime $t(s)/N_0^{0.55}$. Consequently, the cumulative waiting time for the driven part can be expressed as $t(s) \propto N_0^{0.55}s^{1.33}$.

A very good data collapse for the cumulative waiting times for different $N_0$ was found in the whole rigidity range investigated here. The measured cumulative waiting times can be expressed compactly with the scaling functions as:  $t(s,N_0) = N_0^{1.2}\Gamma(s/N_0)$ for $\kappa = 0$ and $t(s,N_0) = N_0^2\Psi(s/N_0^{1.1})$ for $\kappa = 20$. For the semi-rigid polymers the scaling function applies for the driven regime. Similar scaling functions can be found for all intermittent values of $\kappa$. These scaling functions suggest that there may exist a valid scaling hypotheses for the free energy of the confined polymers that would lead to correct ejection dynamics.

\begin{acknowledgments}
Prof.~Jouko Lampinen is thanked for his support of this research. The computational resources of CSC-IT Center for Science, Finland, and Aalto Science-IT project are acknowledged. Part of the work of Joonas Piili has been supported by The Emil Aaltonen Foundation. The work of Pauli Suhonen is supported by The Emil Aaltonen Foundation.
\end{acknowledgments}

\bibliographystyle{ieeetr}
\bibliography{references.bib}

\end{document}